\documentclass[conference]{IEEEtran}
\IEEEoverridecommandlockouts

\usepackage{cite}
\usepackage{amsmath,amssymb,amsfonts}
\usepackage{algorithmic}
\usepackage{graphicx}
\usepackage{textcomp}
\usepackage{xcolor}
\usepackage{booktabs}
\usepackage{makecell}
\usepackage{multirow}
\usepackage{url}
\usepackage{balance}

\providecommand{\Description}[1]{}

\def\BibTeX{{\rm B\kern-.05em{\sc i\kern-.025em b}\kern-.08em
    T\kern-.1667em\lower.7ex\hbox{E}\kern-.125emX}}

\newif\ifSCARFanonymous
\ifdefined\SCARFChinaSysSubmission
  \SCARFanonymoustrue
\fi
\newcommand{\SCARFartifacturl}[1]{%
  \ifSCARFanonymous\textit{[link omitted for anonymous review]}\else\url{#1}\fi}

\begin{document}

\title{SCARF: A Scene-Adaptive Depth-Guided G-3DGS Encoder Accelerator with Semantic Reuse and Fused Dataflow
\ifSCARFanonymous\else
\thanks{This work was supported by the National Key R\&D Program of China (Grant No. 2023YFB4503500), the National Natural Science Foundation of China (Grant No. 62502498), the Beijing Natural Science Foundation (Grant No. L234078), and the Innovation Funding of ICT, CAS (Grant No. E661120).}\fi}

\ifSCARFanonymous
\author{\IEEEauthorblockN{Anonymous Authors}}
\else
\author{
\IEEEauthorblockN{
Zirui Ma\textsuperscript{1,2},
Zhihua Fan\textsuperscript{1,2,*},
Haibin Wu\textsuperscript{1,2},
Wenxing Li\textsuperscript{1,2},
Fulin Zhang\textsuperscript{3,1}, \\
Dongrui Fan\textsuperscript{1},
Xiaochun Ye\textsuperscript{1,2}, and
Wenming Li\textsuperscript{1,2,*}}
\IEEEauthorblockA{\textsuperscript{1}Institute of Computing Technology, Chinese Academy of Sciences, Beijing, China}
\IEEEauthorblockA{\textsuperscript{2}University of Chinese Academy of Sciences, Beijing, China}
\IEEEauthorblockA{\textsuperscript{3}School of Advanced Interdisciplinary Sciences, University of Chinese Academy of Sciences, Beijing, China}
\IEEEauthorblockA{\{mazirui24s, fanzhihua, wuhaibin, liwenxing\}@ict.ac.cn \\
\{fandr, yexiaochun, liwenming\}@ict.ac.cn; zhangfulin20@mails.ucas.ac.cn}
\IEEEauthorblockA{\textsuperscript{*}Corresponding authors: Zhihua Fan and Wenming Li.}
}

\fi

\maketitle

\begin{abstract}
Depth-guided, cost-volume-based Generalizable 3D Gaussian Splatting (G-3DGS) encoders reconstruct scenes as sets of Gaussian primitives directly from multi-view images without requiring per-scene optimization, thereby facilitating feed-forward novel-view synthesis on edge computing platforms. However, their shared encoder architecture presents two primary bottlenecks. First, the per-pixel depth search disregards semantic similarity within the 2D feature space, causing depth search operations to account for over 65\% of memory traffic due to redundant exhaustive candidate scans. Second, the sequential pipeline of depth prediction followed by Gaussian generation neglects local 3D geometric continuity, resulting in more than 52\% of generated Gaussians being redundant prior to instantiation. In this paper, we introduce SCARF, a dataflow-co-designed accelerator that aligns each bottleneck with a corresponding scene-level regularity. To mitigate the first bottleneck, Feature Similarity Depth Reuse reconceptualizes depth search as a problem of semantic locality: depth consistency is determined by affinity in feature space rather than physical address proximity. It transforms exhaustive per-pixel depth searches into anchor-centered local searches. To overcome the second bottleneck, Scene-Adaptive Early Sparsification disrupts the conventional serial depth-to-materialization boundary by probing each tile to infer local scene regularity. It then selectively activates only the minimally sufficient generation pathway, eliminating redundant Gaussian computations before primitive instantiation. On a 27.8 square millimeter 28 nm post-layout design, SCARF delivers a 2.94-fold geometric-mean end-to-end speedup with maximum PSNR degradation below 0.072 percent, 2.23 inferences per joule, and 34.5-fold GPU energy efficiency.
\end{abstract}


\section{Introduction}

\begin{figure}[t]
  \centering
  \includegraphics[width=\linewidth]{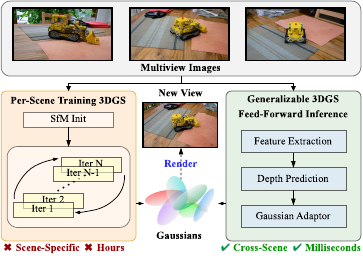}
  \caption{G-3DGS vs.\ per-scene training 3DGS.}
  \Description{Left: per-scene training 3DGS pipeline with SfM initialization and iterative
  gradient optimization (Iter 1 through Iter N), annotated as hours per scene and scene-specific.
  Right: G-3DGS feed-forward pipeline with feature extraction, depth prediction, and Gaussian
  adaptor stages, annotated as milliseconds and cross-scene generalizable. Both paths produce
  a shared Gaussian primitive set that is rendered into a novel view.}
  \label{fig:motivation}
\end{figure}
Latency-sensitive edge applications, including robotic navigation~\cite{huang2024photoslam}, augmented reality~\cite{matsuki2023gaussian-splatting-slam}, and autonomous driving~\cite{yang2023unisim}, require novel-view synthesis that generates scene representations directly from captured images without per-scene optimization~\cite{yu2021pixelnerf,wang2021ibrnet,chen2021mvsnerf}. 3D Gaussian Splatting (3DGS)~\cite{kerbl20233dgs,yu2024mip-splatting,lu2024scaffoldgs} enables real-time rendering through tile-based differentiable rasterization. However, traditional 3DGS is scene-specific and uses Structure-from-Motion initialization~\cite{schonberger2016structure-from-motion-revisited} followed by thousands of gradient iterations to construct the Gaussian set. Consequently, reconstruction takes hours per scene, making traditional 3DGS impractical for feed-forward edge deployment. As illustrated in Fig.~\ref{fig:motivation}, depth-guided, cost-volume-based Generalizable 3D Gaussian Splatting \textbf{(G-3DGS)} encoders~\cite{chen2024mvsplat,xu2024depthsplat,zhang2025transplat} address this limitation by generating Gaussian primitives from multi-view images in one forward pass through cross-view matching, cost-volume depth estimation, and per-pixel regression. This preserves cross-scene generalization without retraining while replacing scene-specific iteration, reducing latency by three to four orders of magnitude relative to scene-specific optimization~\cite{muller2022instant-nngp}.

Despite software progress, these encoders share a common cost-volume dataflow that reveals two critical bottlenecks beyond the reach of software optimization. First, depth search is conducted independently for each pixel, disregarding semantic similarities among pixels. As a result, pixels with similar appearances, and thus nearly identical depth values, still undergo full candidate evaluation. This approach causes the depth search to dominate memory bandwidth usage, accounting for over 65\% of total memory traffic. Second, the sequential pipeline comprising depth prediction and Gaussian generation fails to exploit the continuity inherent in local 3D geometry. Under stringent quality requirements, over 52\% of generated Gaussians are redundant, yet the dataflow cannot eliminate them before generation, causing avoidable computation.

The two bottlenecks stem from two underexplored scene-level regularities: semantically similar pixels tend to share consistent depth estimates, and geometrically continuous surfaces produce redundant Gaussian representations. In this paper, we propose \textbf{SCARF}, a dataflow-co-designed accelerator that exploits semantic similarity during depth search and local geometric continuity during Gaussian materialization to reduce redundant memory accesses and Gaussian computation. SCARF combines \textbf{Feature Similarity Depth Reuse (FSDR)} and \textbf{Scene-Adaptive Early Sparsification (SAES)}. FSDR contracts full depth-candidate scans into semantic-guided local searches, while SAES selects the minimally sufficient generation path before Gaussian materialization. Our principal contributions are as follows:

\textbf{(i)} We introduce \textbf{Feature Similarity Depth Reuse}, which reconceptualizes per-pixel depth search as semantic locality rather than physical proximity. FSDR uses semantic signatures from locality-sensitive hashing (LSH) to index content-addressable memory (CAM). On a cache hit, it replaces the full candidate scan with a localized neighborhood search, reducing cross-view feature-sampling traffic.

\textbf{(ii)} We design \textbf{Scene-Adaptive Early Sparsification}, which disrupts the traditional serial boundary between depth estimation and Gaussian materialization. With a small probe set, SAES evaluates per-tile feature and depth consistency and activates only the minimally sufficient generation path, pruning redundant Gaussian computations while the resulting impact on quality remains controllable.

\textbf{(iii)} We develop \textbf{SCARF}\footnote{The SCARF simulator, Chisel RTL implementation, and evaluation artifact are open source at \SCARFartifacturl{https://github.com/MAdrid1011/SCARF} and archived at \SCARFartifacturl{https://doi.org/10.5281/zenodo.22434931}.}, a \(27.8\,\text{mm}^2\) post-layout accelerator in 28 nm that integrates FSDR and SAES for depth-guided, cost-volume-based G-3DGS encoder inference. Evaluation shows a \(2.94\times\) geometric mean speedup with maximum peak signal-to-noise ratio (PSNR) degradation below 0.072\% and a \(34.5\times\) energy-efficiency improvement over the Jetson Orin NX edge GPU.

\section{Background and Motivation}

\subsection{Generalizable 3DGS Encoders}

Generalizable 3D Gaussian Splatting (G-3DGS) encoders~\cite{chen2024mvsplat,xu2024depthsplat,zhang2025transplat,charatan23pixelsplat,wewer24latentsplat,ye2025nopose} generate scene representations as sets of Gaussian primitives directly from multi-view images in a single feed-forward pass. Existing G-3DGS encoders fall into two categories. Geometry-consistent depth-guided encoders first estimate per-pixel depth through cross-view feature matching and cost-volume aggregation, then generate Gaussian primitives from the resulting depth map. For example, MVSplat~\cite{chen2024mvsplat} uses plane-sweeping cost aggregation to ensure multi-view geometric consistency, enabling robust depth localization even with sparse views. DepthSplat~\cite{xu2024depthsplat} enhances this by incorporating monocular depth priors and self-supervised DINOv2~\cite{caron2021dino,oquab2024dinov2} features, improving robustness in low-texture and occluded regions. TranSplat~\cite{zhang2025transplat} adopts a Transformer-based, depth-aware scheme, integrating cross-view matching~\cite{ding2022transmvsnet,sun2021loftr} and Gaussian parameter prediction within an end-to-end framework. In contrast, direct or implicit encoders, including pixelSplat~\cite{charatan23pixelsplat}, latentSplat~\cite{wewer24latentsplat}, and NoPoSplat~\cite{ye2025nopose}, use probabilistic sampling, latent decoding.

\subsection{Depth-Guided G-3DGS Encoder Pipeline}

Despite differences in backbone and matching strategy, MVSplat, DepthSplat, and TranSplat share a common execution pipeline. A CNN backbone~\cite{he2016deep-residual-learning}, optionally enhanced with Transformer encoder layers~\cite{dosovitskiy2020vit}, processes all input views to generate multi-scale feature maps through hierarchical convolution and self-attention. For each reference-view pixel \((u,v)\), depth estimation is performed independently: each of the \(D\) depth hypotheses \(\{d_k\}\) projects the pixel to the target view, where corresponding features are bilinearly sampled, accumulating per-hypothesis matching costs into a cost volume \(\mathcal{C} \in \mathbb{R}^{H \times W \times D}\). A U-Net~\cite{ronneberger2015unet} refines $\mathcal{C}$ across depth and spatial dimensions, and a Softmax-based depth head converts the refined scores into a continuous depth estimate \(\hat{d}_{u,v} = \sum_k d_k \cdot \mathrm{softmax}(-\mathcal{C}_{u,v})_k\)~\cite{kendall2017endtoendlo}. Finally, the predicted depth map and encoder features are regressed into per-pixel Gaussian attributes, including 3D position $\mu$, covariance $\Sigma$, opacity $\alpha$, and spherical harmonics coefficients $\mathbf{c}_{sh}$, forming the renderable Gaussian set $\mathcal{G}$~\cite{kerbl20233dgs}. The pipeline evaluates $D$ depths for each of the $H\times W$ pixels and emits one primitive per pixel, with no content adaptation. For one scene update and $N$ rendered views, $T_{\mathrm{total}} = T_{\mathrm{enc}} + N T_{\mathrm{rast}}$. Our Orin NX profiling shows that one encoder pass takes about 50$\times$~\cite{charatan23pixelsplat,chen2024mvsplat} as long as one rasterized view, consistent with prior G-3DGS timing reports. The encoder thus remains the dominant cost in edge deployments where the scene representation is refreshed within roughly 50 rendered views.

\subsection{Challenges for Accelerating Depth-Guided G-3DGS Encoders on Edge Platforms}
\label{sec:challenges}

Depth-guided, cost-volume-based G-3DGS encoders exhibit two key characteristics: \textbf{(i)} independent per-pixel depth search across \(D\) hypotheses, and \textbf{(ii)} a serial two-stage dataflow for depth prediction followed by Gaussian generation. We profiled TranSplat~\cite{zhang2025transplat} (Tran), MVSplat~\cite{chen2024mvsplat} (MV), and DepthSplat~\cite{xu2024depthsplat} (Depth) on a Jetson Orin NX, an edge GPU with a limited 2 MB L2 cache, using scenes from the Re10K~\cite{zhou2018stereo}, ACID~\cite{liu2021infinite-nature}, and DL3DV~\cite{ling2024dl3dv} datasets, revealing two challenges to accelerator efficiency.

\paragraph{\textbf{Challenge 1: Independent per-pixel depth search overlooks semantic similarity in 2D feature space, causing redundant memory traffic}}

\begin{figure}[t]
    \centering
    \includegraphics[width=\linewidth]{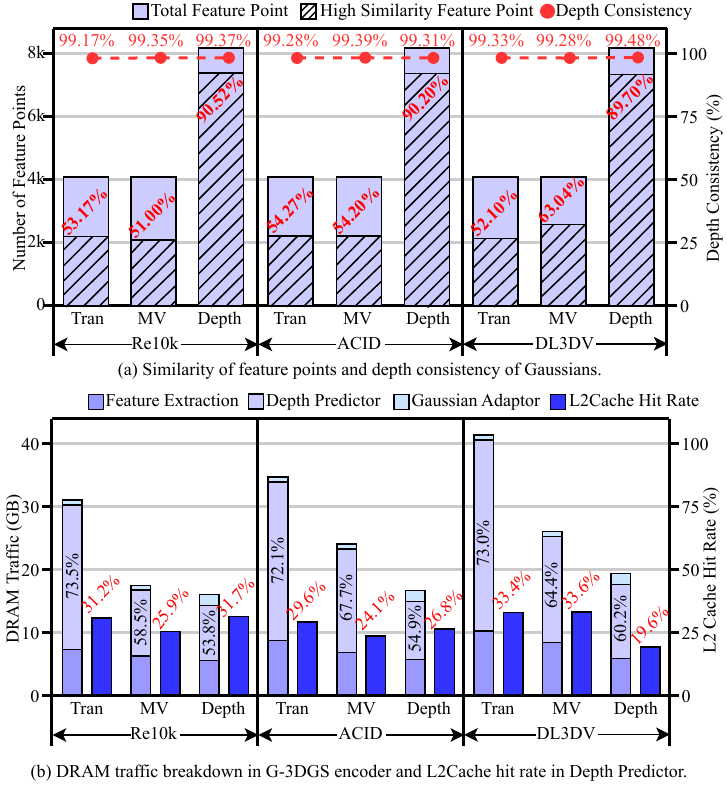}
    \caption{Redundant memory traffic from semantics-oblivious per-pixel depth search with cache-ineffective cross-view sampling.}
    \Description{(a) Feature-similarity statistics: pixel pairs with cosine similarity $>$0.7 account for 51.2\% of all pixels and share 99.5\% depth consistency. (b) Depth search consumes over 65\% of LPDDR5 traffic yet achieves only 31.2\% L2 hit rate, revealing the cache-ineffective access pattern.}
    \label{fig:challenge1}
\end{figure}

Feature maps exhibit a consistent statistical property: pixels with similar feature vectors typically belong to the same object surface and have comparable depth values, based on two priors. First, co-surface pixels share texture and illumination, resulting in highly correlated features extracted by CNNs. Second, object surfaces are geometrically smooth, causing depth to vary continuously within local regions. As shown in Fig.~\ref{fig:challenge1}(a), pixel pairs with cosine feature similarity above 0.7 account for 51.2\% of all pixels on average, with 99.5\% depth consistency. For DepthSplat, which uses self-supervised DINOv2 features, this proportion rises to 91.4\%. These findings indicate that semantically similar pixels provide highly predictive depth estimates.

Existing systems fail to effectively leverage semantic reuse. As shown in Fig.~\ref{fig:challenge1}(b), the depth-search stage generates over 65\% of LPDDR5 memory traffic, despite an average L2 cache hit rate of only 31.2\%. This inefficiency arises from structural issues: cross-view feature sampling across $D$ depth hypotheses produces access patterns that vary with both the depth hypothesis and the target view index, resulting in a lack of physical address locality. Conventional caches depend on address continuity rather than feature-space similarity, causing semantically similar pixels scattered throughout memory to each trigger independent full-candidate scans.

\paragraph{\textbf{Challenge 2: The serial depth-prediction and Gaussian-generation dataflow neglects local continuity in 3D geometry, causing redundant computations}}

\begin{figure}[t]
    \centering
    \includegraphics[width=\linewidth]{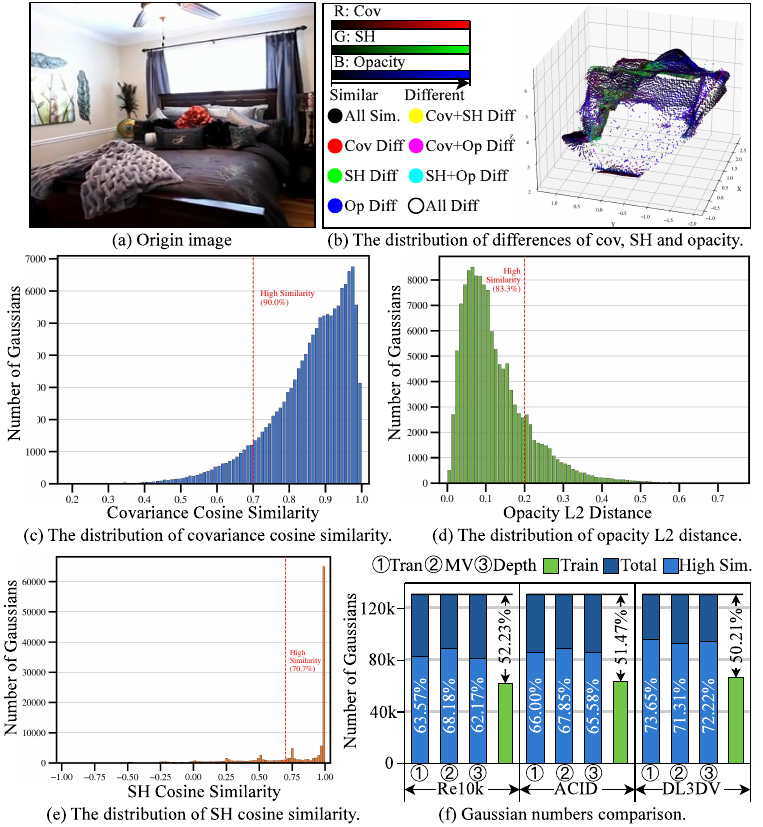}
    \caption{Redundant Gaussian generation from the serial depth-prediction and Gaussian-generation pipeline despite local structural redundancy.}
    \Description{(a)--(e) Covariance, spherical harmonics, and opacity similarity within local 3D Gaussian neighborhoods across models and datasets. (f) Over 52\% of encoder-generated Gaussians are redundant under a 0.1\% quality constraint, yet the serial dataflow cannot detect or skip them before materialization.}
    \label{fig:challenge2}
\end{figure}

Existing G-3DGS encoders employ per-pixel Gaussian generation to maintain generalization, resulting in depth prediction and Gaussian generation accounting for 54\% of the total execution time. To characterize redundancy inside encoder outputs, we analyze local attribute similarity among the generated Gaussian primitives. Fig.~\ref{fig:challenge2}(b) shows the spatial distribution of attribute differences for the scene in Fig.~\ref{fig:challenge2}(a), revealing that neighboring Gaussians are mostly similar. Figs.~\ref{fig:challenge2}(c)--(e) plot the per-attribute similarity distributions and mark the high-similarity boundary used in our redundancy analysis. We count a Gaussian as highly similar on an attribute when the covariance or spherical-harmonic cosine similarity exceeds 0.7, or when the opacity L2 distance is below 0.2. Fig.~\ref{fig:challenge2}(f) further shows that, across the nine model and dataset pairs, 67.8\% of encoder-generated Gaussians on average satisfy all three attribute tests. Thus, adjacent pixels on the same surface produce Gaussian primitives with similar attributes.

Furthermore, using local \(4\times4\) tiles, we merge high-similarity primitives into representative primitives, yielding a 50.9\% Gaussian reduction on average. Across models and datasets, the merged Gaussian sets keep the PSNR loss below 0.1\% in scene rendering, indicating that these high-similarity primitives contribute highly overlapping content to the final rendering. Thus, encoder-generated Gaussian sets contain substantial local structural redundancy. In fact, the fully optimized per-scene 3DGS reference reduces primitives by 51.3\% on average, further showing the large compressible space in scene representation. However, the serial dataflow requires depth prediction for all pixels before Gaussian generation. This prevents depth prediction from observing Gaussian-primitive similarity and causes significant computational redundancy.

\subsection{Our Approach}

\begin{figure*}[t]
    \centering
    \includegraphics[width=\linewidth]{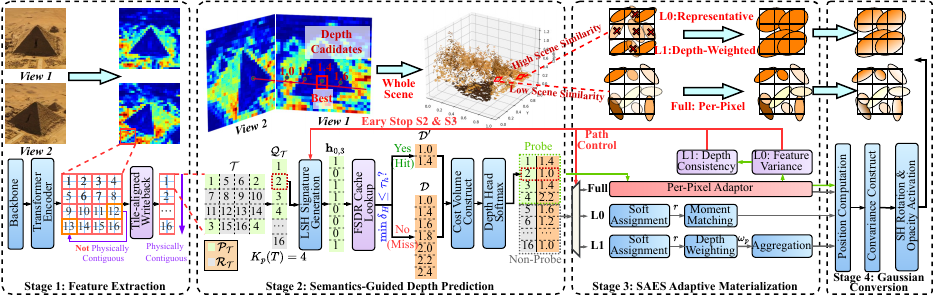}
    \caption{SCARF dataflow overview.}
    \Description{SCARF dataflow overview.}
    \label{fig:dataflow}
\end{figure*}

To address these challenges, we propose SCARF, a dataflow-co-designed accelerator that leverages semantic similarity in 2D feature space and local geometric continuity in 3D space to reduce redundant memory accesses and redundant Gaussian generation:

\paragraph{\textbf{Feature Similarity Depth Reuse (FSDR)}}

To address Challenge 1, we propose FSDR, which replaces address-keyed lookups with semantic-signature-keyed lookups within the depth-search pipeline. Upon a hit, it narrows the candidate set to a cached-anchor neighborhood, replacing the full $D$-candidate scan with a guided local search, thereby eliminating the redundant feature-sampling traffic.

\paragraph{\textbf{Scene-Adaptive Early Sparsification (SAES)}}

To address Challenge 2, we propose SAES, a dataflow that fuses depth prediction with Gaussian materialization. It breaks the strict stage boundary between depth prediction and Gaussian generation. Each tile can activate only the minimally necessary materialization path before full primitive instantiation. By skipping redundant Gaussian computation at an early stage, the proposed approach maintains quality and reduces encoder workload and energy consumption.

\section{SCARF Dataflow}

Building on FSDR and SAES, SCARF restructures the depth-guided G-3DGS encoder pipeline into four stages, as shown in Fig.~\ref{fig:dataflow}. S1 extracts multi-view features, S2 predicts per-pixel depth, S3 materializes Gaussian primitive parameters, and S4 converts network outputs into final Gaussian primitives. Unlike a serialized S2$\rightarrow$S3 pipeline, SCARF fuses S2 and S3 into a tile-level joint pipeline.

\subsection{Stage 1: Feature Extraction}

S1 extracts backbone features from all input views by sequentially applying CNN convolutional layers and Transformer encoder layers to produce multi-scale feature maps. 

\paragraph{\textbf{Model-agnostic runtime configuration}}

Depth-guided G-3DGS encoders vary in convolutional and Transformer layer counts and may include a Vision Transformer (ViT) branch. Memory-mapped input/output (MMIO) configuration registers manage these differences uniformly. This dataflow eliminates model-specific branches and supports depth-guided topologies on unified hardware.

\paragraph{\textbf{Tile-aligned writeback layout}}

Feature maps at all scales are divided into fixed-size tiles before off-chip writeback. A linear layout keeps data contiguous within each tile and stores tiles sequentially, placing cross-view features of the same tile adjacent in memory. The layout matches the probe-prioritized execution order of S2, and the critical pixel reads therefore touch only a few consecutive cache lines and avoid data reordering or inter-block movement before S2.

\subsection{Stage 2: Semantics-Guided Depth Prediction}

S2 uses FSDR-guided candidate contraction to reduce memory traffic during per-pixel cross-view depth prediction.

\paragraph{\textbf{Computation order scheduling}}

For an input tile $\mathcal{T}$, S2 constructs an ordered pixel queue $\mathcal{Q}_{\mathcal{T}}=[\mathcal{P}_{\mathcal{T}},\ \mathcal{R}_{\mathcal{T}}]$, where $\mathcal{P}_{\mathcal{T}}$ denotes the probe set used by S3 and $\mathcal{R}_{\mathcal{T}}=\mathcal{T}\setminus\mathcal{P}_{\mathcal{T}}$ contains the remaining pixels in row-major order. S2 schedules probe pixels first and places corner probes at the front of $\mathcal{P}_{\mathcal{T}}$. This order establishes local evidence for SAES before issuing the remaining depth searches.

\paragraph{\textbf{Depth-candidate contraction}}

For pixel $(u,v)$, S2 performs semantics-guided selection over $\mathcal{D}=\{d_i\}_{i=1}^{D}$. Given feature $\mathbf{x}_{u,v}\in\mathbb{R}^{C}$ and hyperplanes $\{\mathbf{r}_k\}_{k=1}^{K}$, S2 computes the LSH~\cite{charikar2002similarity-estimation} signature as
\[
h_k(\mathbf{x}_{u,v})=\mathbb{I}\!\left(\mathbf{r}_k^\top \mathbf{x}_{u,v}\ge 0\right),\qquad
\mathbf{h}_{u,v}=[h_1,\ldots,h_K].
\]

The signature $\mathbf{h}_{u,v}$ is queried against the FSDR cache, where the Hamming distance is calculated as
\[
\delta_H(\mathbf{h}_{u,v},\mathbf{h}_j)=
\sum_{k=1}^{K}\left|h_k(\mathbf{x}_{u,v})-h_k(\mathbf{x}_j)\right|.
\]

Each bit records which side of a random hyperplane the feature vector falls on. The Hamming distance therefore counts hyperplane boundary crossings between two vectors and hardware-efficiently approximates their angular distance in the feature space.

If $\min_j \delta_H(\mathbf{h}_{u,v},\mathbf{h}_j)\le \tau_h$, FSDR retrieves the cached depth $d_c$ and forms the local candidate set $\mathcal{D}^{\prime}$ by selecting the nearest $D/R$ depth values around $d_c$ from the global candidate set $\mathcal{D}$. Otherwise, SCARF uses the global candidate set $\mathcal{D}$. This contraction occurs only before cost-volume construction. The retained candidates still use the original global depth values and pass through the same cross-view feature sampling, cost-volume construction, pretrained U-Net refinement, and depth-head regression. Thus, hit pixels preserve the baseline depth-estimation path, except that the cost-volume support is restricted around the cached depth. Weak semantic matches fall back to full global search. The final estimate $\hat d_{u,v}$ is written back with the current feature signature into the FSDR cache for reuse by subsequent pixels.



\subsection{Stage 3: SAES Adaptive Materialization}

S3 receives depth outputs from S2 and materializes Gaussian primitive attributes through the SAES three-level path hierarchy. SAES makes a tile-level path decision, but it does not replace the pretrained Gaussian adaptor. Probe pixels always use their original features and predicted depths as adaptor inputs, producing probe Gaussians through the same baseline adaptor. The decision only controls non-probe pixels. L0 and L1, corresponding to the representative and lightweight paths, approximate them from the probe Gaussians in Gaussian-parameter space, while Full keeps per-pixel adaptor execution. The hierarchy follows a first-hit progression \textbf{L0}$\rightarrow$\textbf{L1}$\rightarrow$\textbf{Full}, where each level attempts the most aggressive pruning before falling through to the next.

\paragraph{\textbf{Probe-set construction}}

Let the current tile be $\mathcal{T}$ with side length $T$ pixels. The ordered probe set \(\mathcal{P}_{\mathcal{T}} \subset \mathcal{T}\) starts from the four corners and expands slowly as tile size grows. To balance representativeness and overhead, probes are selected using a \emph{corner-anchored and slowly expanding} strategy. For \(T=4\), only the four corners are used. For \(T>4\), the number of probes is defined as
\[
K_p(T)=4+\left\lceil 2\log_2\left(\frac{T}{4}\right)\right\rceil,\qquad
|\mathcal{P}_{\mathcal{T}}|=K_p(T).
\]

Let \( M = K_p(T) - 4 \) be the number of newly added probes. The area, excluding the four corners, is divided into a $G\times G$ subgrid, where \( G = \lceil \sqrt{M} \rceil \). From each subgrid, the pixel closest to the center is selected. This method prevents decision bias caused by probe clustering and ensures that the number of probes increases gradually with the tile size.

\paragraph{\textbf{L0: feature-consistency decision}}

The L0 stage makes the earliest decision based on feature consistency among probes. Let $\bar{\mathbf{f}}$ and $\sigma_{\mathrm{feat}}^2(\mathcal{P}_{\mathcal{T}})$ denote the mean and variance of the probe features $\{\mathbf{f}_p\}_{p\in\mathcal{P}_{\mathcal{T}}}$. If $\sigma_{\mathrm{feat}}^2(\mathcal{P}_{\mathcal{T}})<\tau_f$, the tile is classified as L0 and routed to the representative path. Low feature variance among probes indicates that the tile lies on a single-surface region. Such tiles produce Gaussians with nearly identical attributes, making probe-level representation sufficient. Only representative Gaussians from the probe path are retained. Non-probe pixels are merged into their corresponding probes via a scene-adaptive soft assignment~\cite{comaniciu2002meanshift}. For any pixel $i\in\mathcal{T}$, the assignment weight $r_{i\rightarrow p}^{(\mathrm{L0})}$ combines spatial proximity and feature affinity through a joint bilateral kernel, with the spatial bandwidth modulated by the tile's own feature variance $\sigma_{\mathrm{feat}}^2$. In smooth tiles, the spatial term defers to feature similarity. In textured tiles, assignments are tightened and localized by the spatial term:
\[
r_{i\rightarrow p}^{(\mathrm{L0})}
=
\frac{
\exp\!\left(
-\frac{\sigma_{\mathrm{feat}}^2\cdot\|\mathbf{x}_i-\mathbf{x}_p\|_2^2}{\beta_x^2}
-\frac{\|\mathbf{f}_i-\mathbf{f}_p\|_2^2}{\beta_f^2}
\right)
}{
\sum_{q\in\mathcal{P}_{\mathcal{T}}}
\exp\!\left(
-\frac{\sigma_{\mathrm{feat}}^2\cdot\|\mathbf{x}_i-\mathbf{x}_q\|_2^2}{\beta_x^2}
-\frac{\|\mathbf{f}_i-\mathbf{f}_q\|_2^2}{\beta_f^2}
\right)
}.
\]

Using $r_{i\rightarrow p}^{(\mathrm{L0})}$, each probe's representative Gaussian is updated as the weighted mean $\mu_p'$ and covariance $\Sigma_p'$ via first- and second-moment matching~\cite{fan2024lightgaussian}. This keeps the adaptor-generated probe attributes as anchors and changes only how locally redundant pixels are absorbed. Spherical harmonics and opacity are averaged under range constraints. The L0 path offers the highest compression and is ideal for areas with smooth textures and geometry.

\paragraph{\textbf{L1: depth-consistency decision}}

L1 is accessed only if L0 misses. At this stage, lightweight depth estimation is performed on the probes. Let $\bar{d}$ and $\sigma_d(\mathcal{P}_{\mathcal{T}})$ denote the mean and standard deviation of probe depths $\{d_p\}_{p\in\mathcal{P}_{\mathcal{T}}}$. If $\sigma_d(\mathcal{P}_{\mathcal{T}})<\tau_d$, the tile is classified as L1 and directed to the lightweight path. A low depth standard deviation indicates a locally planar surface. On such surfaces, Gaussian positions and covariances vary minimally across pixels, so probe-anchored aggregation captures the dominant geometry without per-pixel instantiation. L1 retains the probe-constrained aggregation framework of L0 but introduces a depth reliability factor $\omega_p$ to recalibrate aggregation weights. This factor normalizes each probe's depth deviation $|d_p-\bar{d}|$ by the tile's own depth spread $\sigma_d(\mathcal{P}_\mathcal{T})$. Near-planar tiles tolerate little deviation, whereas depth-varying tiles have proportionally greater tolerance for deviation:
\[
\omega_p=
\exp\!\left(-\frac{|d_p-\bar d|}{\beta_d\cdot\sigma_d(\mathcal{P}_{\mathcal{T}})+\epsilon}\right).
\]

Because L1 passes depth validation, it applies less compression than L0. This makes L1 better suited for areas characterized by smooth geometry.

\paragraph{\textbf{Full: complete computation}}

If L1 does not hit, the tile is assigned to the Full path, in which complete S2 depth search and per-pixel Gaussian adaptor execution in S3 are performed. The Full path is intended for geometrically complex regions, such as object boundaries and depth discontinuities, where preserving all primitives is necessary for rendering quality. This fallback keeps the baseline per-pixel adaptor path for tiles whose local evidence supports neither representative materialization nor lightweight materialization.

\subsection{Stage 4: Gaussian Conversion}

S4 uses the Gaussian Generation Unit (GGU) Array to convert retained depth and attributes into renderable primitives $\{\mu,\Sigma,\alpha,\mathbf{c}_{sh}\}$, then commits them to off-chip memory through two bandwidth optimizations.

\paragraph{\textbf{Compact writeback format}}

S4 reduces output bandwidth by using a compact writeback format aligned with the Gaussian representation. It writes only the six independent upper-triangular elements of the symmetric covariance matrix and stores only the valid spherical-harmonics coefficients for the configured order. This design cuts writeback bytes per Gaussian by over 35\% without altering computational semantics and ensures that output bandwidth exhibits linear scaling with scene complexity.

\paragraph{\textbf{Streaming commit}}

During S4 execution, results are committed to the writeback buffer in tile-local Gaussian order, avoiding the burstiness and queuing overhead of conventional batch-based writeback. For tiles routed to the L0 representative or L1 lightweight path chosen by SAES, S4 receives only the retained outputs from that path, thereby further reducing conversion and writeback traffic.

\begin{figure*}[t]
    \centering
    \includegraphics[width=\linewidth]{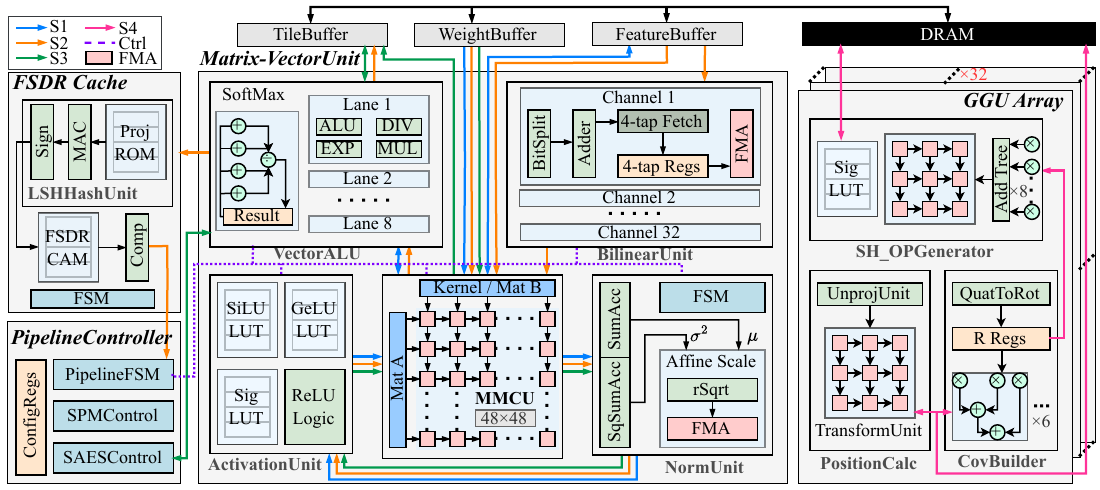}
    \caption{SCARF architecture overview.}
    \Description{Five tightly coupled components: MVU (centered on a 48$\times$48 MMCU systolic array), FSDR Cache, GGU Array, control subsystem, and SPM subsystem, organized to keep high-frequency feature--depth--Gaussian dependencies within an on-chip closed loop.}
    \label{fig:arch}
\end{figure*}

\section{SCARF Architecture}

The SCARF architecture (Fig.~\ref{fig:arch}) comprises five interconnected components: the Matrix Vector Unit (MVU), FSDR Cache, GGU Array, control subsystem, and scratchpad memory (SPM) subsystem. The MVU shares a matrix-multiplication compute unit (MMCU) across S1 through S3, amortizing compute area across their diverse operators. The FSDR Cache intercepts per-pixel depth-search requests at the S2 front end, eliminating redundant candidate scans before they reach the MMCU. The GGU Array offloads geometric conversion from the MVU, allowing computation of the next tile while primitive construction proceeds concurrently. Together, these units keep the high-frequency dependency chain among features, depths, and Gaussians on chip, leaving only essential inputs and outputs on the DRAM path. 

\subsection{Matrix-Vector Unit}

Across S1, S2, and S3, the operator mix shifts from matrix-intensive convolution and attention in S1, through irregular bilinear sampling and reduction in S2, to SAES-specific vector statistics in S3. Provisioning stage-specific arrays would fragment area and force costly data reordering at stage boundaries. SCARF instead organizes the MVU around a shared MMCU for convolution, general matrix multiplication (GEMM), and attention, augmented by a BilinearUnit for irregular cross-view sampling, a VectorALU for reductions and SAES path-selection statistics, and fused NormUnit and ActivationUnit to eliminate intermediate writebacks. 

\paragraph{\textbf{S1: feature extraction}}

The MMCU operates in Conv, GEMM, and Attention modes. In Conv mode, convolutions are mapped via \textit{im2col} with fused normalization and activation. The NormUnit and ActivationUnit handle these operations to eliminate intermediate writeback. In GEMM mode, the same array processes QKV projections and FFN blocks. In Attention mode, the VectorALU executes scaled dot-product and normalization alongside the MMCU, enabling self-attention and cross-attention.

\paragraph{\textbf{S2: depth prediction}}

S2 reuses the MMCU across both Conv and GEMM modes for U-Net refinement and depth-head regression, switching between them via weight-switch commands to the WeightBuffer to avoid stalls caused by array reconfiguration. A dedicated 32-way BilinearUnit absorbs this irregular traffic concurrently, keeping the MMCU fully occupied with refinement. To produce $\hat d_{u,v}$, the VectorALU then normalizes over depth candidates and computes the expectation over those candidates.

\paragraph{\textbf{S3: feedback execution and Gaussian-parameter prediction}}

The single-instruction, multiple-data (SIMD) datapath of the VectorALU computes the L0 feature variance, \(\sigma_{\mathrm{feat}}^2(\mathcal{P}_{\mathcal{T}})\), and L1 depth standard deviation, \(\sigma_d(\mathcal{P}_{\mathcal{T}})\), online through a reduction tree with constant path-selection latency for any probe count. In the representative-generation path, the VectorALU performs weighted moment matching to incorporate non-probe pixels into representative Gaussians. In the Full path, the MMCU Conv mode is reused for parameter prediction with fused NormUnit and ActivationUnit execution.

\subsection{Feature Similarity Depth Reuse Cache}

\paragraph{\textbf{Storage structure}}

\begin{figure}[t]
    \centering
    \includegraphics[width=\linewidth]{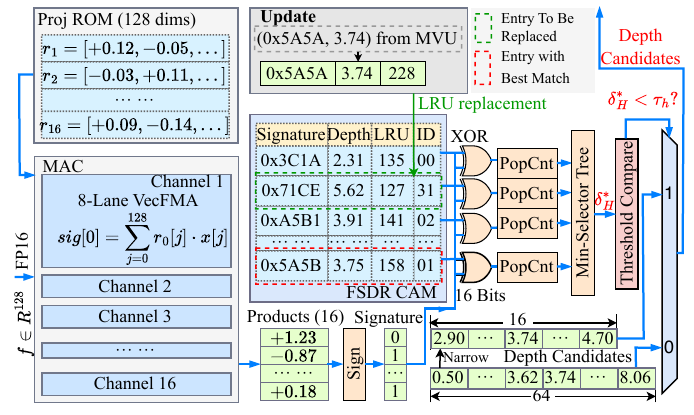}
    \caption{FSDR cache datapath.}
    \Description{LSHHashUnit compresses the feature vector to a 16-bit semantic signature. A 32-entry CAM array performs Hamming-distance matching. On a hit the FSDR Controller contracts the candidate set around the cached depth anchor. On a miss the full set is retained.}
    \label{fig:fsdrcache}
\end{figure}

As illustrated in Fig.~\ref{fig:fsdrcache}, the FSDR Cache uses a 32-entry CAM because its primary reuse targets, feature-similar pixels, are physically scattered across the frame. Pixels with nearly identical features and depths may have distant memory addresses, so address-domain caches miss most reuse opportunities regardless of capacity. CAM instead indexes signature values through parallel associative lookup across all entries, locating semantic matches regardless of physical address. Each 41-bit entry stores a valid bit, a 16-bit LSH signature, an FP16 depth anchor, and an 8-bit least-recently-used (LRU) counter.

\paragraph{\textbf{Key generation and lookup}}

The LSHHashUnit projects the 128-dimensional feature vector $\mathbf{x}_{u,v}$ onto $K=16$ random hyperplanes stored in on-chip ROM, extracting sign bits to generate the signature $\mathbf{h}_{u,v}$. Random hyperplane projection preserves the cosine-similarity structure of CNN features: pixels lying on the same surface cluster into nearby regions of signature space, so a small Hamming distance reliably predicts depth agreement. For each cache entry $j$, the following expression computes the distance:
\[
\delta_H^{(j)}=\mathrm{popcount}(\mathbf{h}_{u,v}\oplus\mathbf{h}_j).
\]

A 32-entry parallel XOR-popcount array evaluates all entries in one cycle. A reduction tree then selects the best-matching entry $j^*$ and its distance $\delta_H^*$, and the cached depth $d_c$ is read and compared to $\tau_h$ in the next cycle. Upon a hit, the Controller narrows the candidate set from $\mathcal{D}$ to $\mathcal{D}'$ centered at $d_c$, with size \(|\mathcal{D}'| = D/R\), so that the search focuses on the most likely depth neighborhood.

\paragraph{\textbf{Update policy}}

After the current pixel completes regression and produces $\hat d_{u,v}$, the Controller updates the semantic-neighborhood cache by writing back \((\mathbf{h}_{u,v}, \hat{d}_{u,v})\). Empty entries are filled first. Once full, the entry with the smallest LRU counter is replaced to prioritize active semantic patterns in the current frame. 

\subsection{Gaussian Generation Unit}

The GGU Array performs the final geometric mapping, converting raw per-pixel network outputs into renderable Gaussian primitives. Since these geometric operations are independent across Gaussians, the array comprises 32 processing elements, each executing a three-stage pipeline: \textit{PositionCalc} back-projects and applies extrinsic transformations to obtain the world-space position $\mu$. \textit{CovBuilder} expands the quaternion into a rotation matrix $R$ and constructs the covariance matrix \(\Sigma = R S S^\top R^\top\). \textit{SH\_OPGenerator} rotates SH coefficients into world coordinates via per-band Wigner-D multiplication and applies sigmoid activation to opacity.

\paragraph{\textbf{Rotation-matrix bypass}}

Within a GGU processing element, spherical harmonics rotation is the most memory-intensive sub-stage, comprising about 43\% of total GGU memory traffic. Since the required rotation matrix $R$ is mathematically identical to the one computed in CovBuilder, SCARF directly connects CovBuilder's R-Regs to SH\_OPGenerator's input via a bypass. This eliminates the downstream stall while waiting for them to be reloaded before Wigner-D multiplication, enabling SH\_OPGenerator to start immediately after CovBuilder has completed its work.

\paragraph{\textbf{Precision-sensitive mixed-precision strategy}}

3DGS rendering is sensitive to geometric precision: multi-stage accumulation in back-projection and covariance construction can amplify FP16 errors into visible artifacts, whereas spherical harmonics rotation is more tolerant of reduced precision. SCARF therefore applies FP32 selectively to geometry-critical steps, including PositionCalc, CovBuilder, and the rotation matrices in SH\_OPGenerator, preserving geometric fidelity without the bandwidth cost of full FP32.

\begin{figure}[t]
    \centering
    \includegraphics[width=240pt]{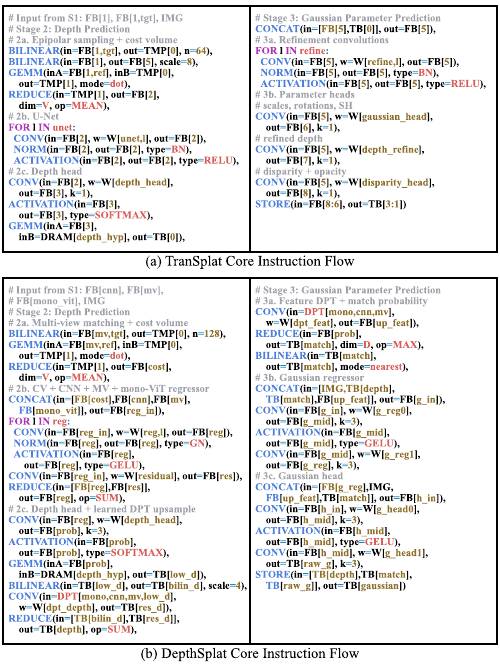}
    \caption{S2/S3 instructions for TranSplat and DepthSplat.}
    \Description{Pseudocode listings of SCARF primitive instructions for TranSplat and DepthSplat. Each panel places S2 on the left and S3 on the right. The TranSplat sequence shows epipolar sampling, cost-volume refinement, depth regression, and Gaussian parameter heads. The DepthSplat sequence shows the additional consumption of mono ViT features and DPT upsampling paths while using the same primitive instruction set.}
    \label{fig:schedule}
\end{figure}

\subsection{SCARF Scheduling and Model Adaptation}

The algorithmic evolution of depth-guided, cost-volume-based G-3DGS encoders outpaces hardware iteration cycles. SCARF bridges this gap with a programmable architecture that enables a single hardware instance to efficiently support multiple models within the same depth-guided, cost-volume-based encoder family. Its programming model uses instruction sequences to define the computation graph, with each instruction specifying the operator type, input/output buffer addresses, and hardware configuration. At runtime, the PipelineFSM reads instructions sequentially, dispatches tasks to hardware units based on the operator type, and manages data movement among the buffers.

SCARF defines six hardware primitives: \texttt{CONV} for convolution and deconvolution, \texttt{GEMM} for matrix multiplication, \texttt{NORM} for BatchNorm and LayerNorm, \texttt{ACTIVATION} for ReLU, GELU, and Softmax, \texttt{BILINEAR} for bilinear sampling, and \texttt{REDUCE} for Mean, Max, and Sum. Higher-level operators such as Transformer attention and epipolar sampling are decomposed into these primitives. This abstraction maintains a stable hardware base while adapting to model variations through instruction scheduling. Fig.~\ref{fig:schedule} compares the S2 and S3 instructions of TranSplat and DepthSplat. The comparison exposes two concrete schedule differences. First, DepthSplat carries the mono ViT feature branch from S1 into S2, where cost-volume refinement fuses CNN, multi-view, and mono features, while TranSplat refines the epipolar cost volume and regresses depth. Second, DepthSplat adds DPT-based upsampling before Gaussian prediction, whereas the lighter depth and feature outputs go directly into the Gaussian heads that belong to TranSplat.

\section{Methodology}

\paragraph{\textbf{Experimental Platform}}

We implemented the SCARF RTL design using the TSMC 28 nm process and the OpenROAD physical-design flow, including synthesis, placement, routing, and post-layout timing analysis~\cite{ajayi2019openroad}. On-chip SRAM blocks were generated using a commercial memory compiler within the same technology node. The design operates at 1 GHz and features a fully pipelined architecture. To evaluate SCARF under realistic workloads, we developed a cycle-accurate simulator that models the SCARF microarchitecture and integrates Ramulator~\cite{luo2023ramulator2} to capture detailed DRAM timing for Micron LPDDR4X-4266. The simulator performs both functional and timing simulations, accepts multi-view images as input, and outputs predicted 3D Gaussian parameters. Performance metrics are derived from simulator cycle counts. Logic power is estimated from the post-layout gate-level netlist, TSMC 28 nm Liberty timing and power files, and workload switching activity generated by the cycle-accurate simulator. The switching activity is annotated in the OpenROAD and OpenSTA analysis flow to estimate dynamic power and leakage power. On-chip SRAM energy is estimated from simulator access counts and memory-compiler energy numbers. Using DRAM activity traces generated by the cycle-accurate simulator, DRAMPower~\cite{steiner2025drampower5} provides the corresponding off-chip memory energy estimate.

\paragraph{\textbf{Workloads}}

We selected three representative depth-guided, cost-volume-based G-3DGS models: MVSplat~\cite{chen2024mvsplat} (MV), TranSplat~\cite{zhang2025transplat} (Tran), and DepthSplat~\cite{xu2024depthsplat} (Depth), which differ in architecture and computational characteristics. MVSplat is the lightest, TranSplat has medium complexity, and DepthSplat is the most compute-intensive due to its heavier ViT backbone and larger depth search space. The datasets used include Re10K~\cite{zhou2018stereo}, ACID~\cite{liu2021infinite-nature}, and DL3DV~\cite{ling2024dl3dv}, covering diverse indoor/outdoor, driving, and dense multi-view scenes. All models use official pretrained weights.



\paragraph{\textbf{Baseline}}

Previous accelerators have primarily focused on per-scene NeRF rendering~\cite{noh2025flexnerfer, li2023instant-3d} and 3DGS rasterization~\cite{pei2025gcc, lee2024gscore}, but none have targeted feed-forward G-3DGS encoder construction. We use the Jetson Orin NX 16 GB as the system-level baseline, a common edge AI platform in robotics and autonomous driving. It features an Ampere GPU with 1,024 CUDA cores, 16 GB of LPDDR5 memory at 102.4 GB/s, and a 2 MB L2 cache, running at 918 MHz in MAXN mode (25 W TDP). Encoder and rasterizer latencies were measured separately using Nsight Systems. The optimized software stack employs PyTorch~\cite{paszke2019pytorch} with cuDNN~\cite{chetlur2014cudnn}, cuBLAS, optimized epipolar sampling, cost-volume and attention paths, and custom CUDA kernels for model-specific operators. The implementation uses FP16 precision and CUDA Graphs throughout.

\section{Evaluation}

\subsection{Performance and Quality}

\begin{figure}[t]
    \centering
    \includegraphics[width=\linewidth]{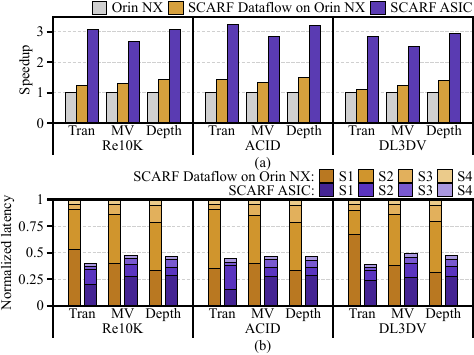}
    \caption{End-to-end speedup of SCARF.}
    \Description{Two-panel chart showing end-to-end speedup across three execution modes and normalized S1 to S4 latency breakdowns for SCARF Dataflow on Orin NX and SCARF ASIC.}
    \label{fig:performance}
\end{figure}
\begin{figure}[t]
    \centering
    \includegraphics[width=\linewidth]{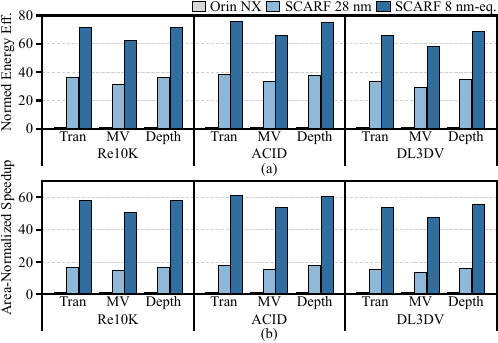}
    \caption{Area efficiency and energy efficiency of SCARF.}
    \Description{Two-panel bar charts showing normalized energy efficiency and area-normalized speedup of Orin NX, measured SCARF at 28 nm, and estimated SCARF at 8 nm-equivalent across multiple model and dataset combinations.}
    \label{fig:efficiency}
\end{figure}

\paragraph{\textbf{End-to-End Performance}}

Fig.~\ref{fig:performance}(a) summarizes the end-to-end performance of SCARF ASIC over Orin NX. Across nine model and dataset pairs, SCARF reaches a geometric-mean speedup of 2.94$\times$ and an area-normalized throughput of 0.168 inf/s/mm$^2$, which is 16.1$\times$ higher than Orin NX. The gain comes from algorithm-level work reduction and hardware mapping. In S2, FSDR uses semantic hashing and cache lookup to select a local depth window, reducing depth-candidate search and feature-buffer accesses. At the tile level, SAES identifies local redundancy and routes selected tiles to the representative or lightweight path, reducing full Gaussian materialization. To separate the contribution of the dataflow from ASIC mapping, we implement SCARF Dataflow on Orin NX. This configuration keeps the Orin NX GPU backend unchanged, adds FSDR depth-window selection and SAES tile-level path selection to the encoder. It reaches a geometric-mean speedup of 1.33$\times$, showing that FSDR and SAES also reduce useful work on the GPU. The GPU gain remains smaller because hashing, path selection, compaction, and small-kernel scheduling lower GPU efficiency.

\paragraph{\textbf{Latency Breakdown}}

Fig.~\ref{fig:performance}(b) decomposes the latency of SCARF Dataflow on Orin NX and SCARF ASIC into S1 through S4. The dataflow-on-GPU stack is normalized to 1 for each model and dataset pair. The ASIC stack uses the same baseline, so each colored segment shows the remaining latency of the corresponding stage after hardware mapping. The ASIC stacks occupy 0.39 to 0.49 of the normalized dataflow height. S2 contributes the largest shrinkage. It reaches a geometric-mean speedup of 3.96$\times$ because FSDR contracts the depth-search space from \(D\) to \(D/4\) and reduces feature-buffer accesses. S3 reaches a geometric-mean speedup of 1.80$\times$ because SAES reduces tiles that require full per-pixel Gaussian materialization. S1 and S4 reach geometric-mean speedups of 1.62$\times$ and 1.54$\times$. S1 improves because its convolution and attention layers map to long dense MMCU kernels with fused normalization and activation, avoiding small-kernel launch and intermediate-writeback overheads in the GPU baseline. S4 improves because the GGU maps per-Gaussian conversion operations onto a dedicated pipeline, while SAES reduces the number of Gaussians entering S4.

\paragraph{\textbf{Energy Efficiency}}

Figure~\ref{fig:efficiency} compares the measured 28\,nm results of SCARF with 8\,nm-equivalent results estimated by DeepScaleTool~\cite{sarangi2021deepscaletool}. The 8\,nm-equivalent results control for the process-node gap between SCARF and Orin NX, so they primarily reflect the contribution of architecture and dataflow. Across multiple model and dataset pairs, SCARF reaches a geometric-mean energy-efficiency improvement of 68.2$\times$ and an area-normalized throughput improvement of 55.2$\times$ under this process-normalized view. This gain comes from work reduction and specialized hardware mapping. FSDR and SAES reduce depth search, feature-buffer traffic, and Gaussian materialization, so each inference performs fewer high-energy accesses and redundant computations. Under the less advanced 28\,nm process, the measured post-layout implementation still reaches 34.5$\times$ and 16.1$\times$. Concretely, the process gap reduces the energy-efficiency gain by 1.98$\times$ on average and the area-normalized throughput gain by 3.43$\times$ on average. Therefore, the architecture and dataflow design of the G-3DGS encoder account mainly for the efficiency advantage achieved by the SCARF design.

\paragraph{\textbf{Overall Quality}}

Table~\ref{tab:quality} shows that SCARF stays close to the baseline across all nine pairs. The maximum changes are 0.072\% in PSNR, 0.24\% in SSIM, and 0.002 in LPIPS, with no consistent degradation. Quality is preserved by the adaptive mechanisms in both approximation stages. FSDR's local candidate window bounds depth deviations and confines DL3DV fluctuations to ${\leq}$0.19\% PSNR at abrupt depth boundaries. SAES's scene-adaptive $r_{i\rightarrow p}$ and $\omega_p$ suppress over-smoothing at textured boundaries and relax in uniform regions. For example, TranSplat on ACID gains 0.07\% PSNR and $-$0.001 LPIPS. Where metrics diverge, the shifts reflect localized artifacts rather than global fidelity loss.

\begin{table}[t]
    \centering
    \caption{SCARF rendering quality versus baseline.}
    \label{tab:quality}
    \resizebox{\columnwidth}{!}{
    \begin{tabular}{ccccc}
        \toprule
        \textbf{Model} & \textbf{Dataset} &
        \makecell{\textbf{PSNR}$\uparrow$\\\small\textbf{Base~/~SCARF}} &
        \makecell{\textbf{SSIM}$\uparrow$\\\small\textbf{Base~/~SCARF}} &
        \makecell{\textbf{LPIPS}$\downarrow$\\\small\textbf{Base~/~SCARF}} \\
        \midrule
        \multirow{3}{*}{Tran}  & Re10K & 28.08~/~28.07 & 0.919~/~0.917 & 0.128~/~0.130 \\
                                    & ACID  & 29.58~/~29.60 & 0.899~/~0.901 & 0.145~/~0.144 \\
                                    & DL3DV & 25.08~/~25.07 & 0.854~/~0.852 & 0.178~/~0.180 \\
        \cmidrule(l){2-5}
        \multirow{3}{*}{MV}    & Re10K & 30.22~/~30.23 & 0.942~/~0.944 & 0.128~/~0.123 \\
                                    & ACID  & 32.08~/~32.06 & 0.916~/~0.915 & 0.144~/~0.146 \\
                                    & DL3DV & 28.02~/~28.05 & 0.914~/~0.912 & 0.145~/~0.146 \\
        \cmidrule(l){2-5}
        \multirow{3}{*}{Depth} & Re10K & 28.43~/~28.42 & 0.925~/~0.923 & 0.122~/~0.123 \\
                                    & ACID  & 29.43~/~29.44 & 0.910~/~0.910 & 0.137~/~0.137 \\
                                    & DL3DV & 26.93~/~26.98 & 0.885~/~0.886 & 0.157~/~0.156 \\
        \bottomrule
    \end{tabular}}
\end{table}

\begin{figure}[t]
    \centering
    \includegraphics[width=\linewidth]{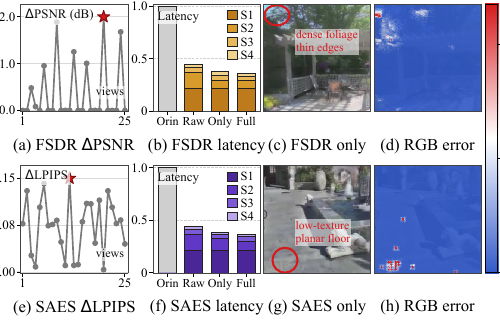}
    \caption{Worst-case view losses, latency and error maps.}
    \Description{Two rows showing FSDR and SAES worst cases. Each row includes a view-wise quality-loss trace, a latency breakdown, the mechanism output, and an RGB error map.}
    \label{fig:quality_error}
\end{figure}

\paragraph{\textbf{Worst-Case Error Analysis}}

Fig.~\ref{fig:quality_error} shows view-wise quality losses, normalized latency breakdowns, and RGB error maps for the worst FSDR and SAES views. Fig.~\ref{fig:quality_error}(a) reports FSDR-only \(\Delta\)PSNR across all evaluated views, and Fig.~\ref{fig:quality_error}(e) reports SAES-only \(\Delta\)LPIPS. The red stars mark the worst views shown in Fig.~\ref{fig:quality_error}(c) and (g). For FSDR, the error remains localized because the full-candidate fallback avoids narrowing depth search when semantic matching is unreliable. The remaining gain comes from restricting reliable S2 matches to local depth windows and reducing feature-buffer accesses. For SAES, the worst LPIPS case concentrates errors at sparse local structures where representative or lightweight materialization changes Gaussian attributes. These local errors are more visible to LPIPS because LPIPS is sensitive to boundaries, texture, and shadow. In Table~\ref{tab:quality}, the largest LPIPS increase is only from 0.178 to 0.180, indicating no persistent degradation in global quality from the local worst-view errors.

\subsection{Source of Performance Gains}

\paragraph{\textbf{Speedup Ablation}}

\begin{figure}[t]
    \centering
    \includegraphics[width=\linewidth]{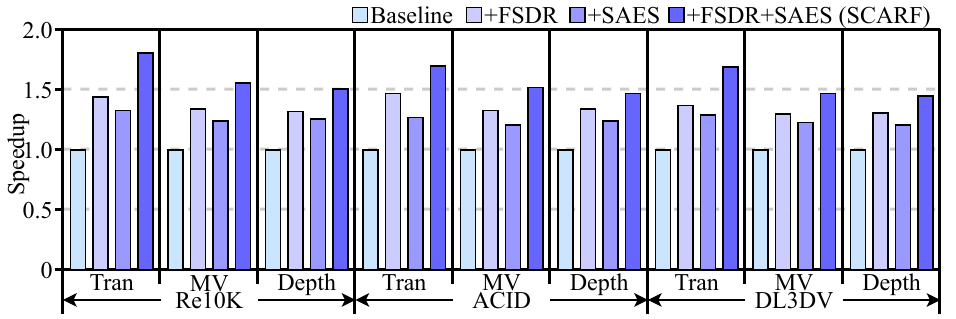}
    \caption{Ablation study of FSDR and SAES.}
    \Description{Grouped bar chart showing speedup of FSDR only, SAES only, and FSDR+SAES relative to SCARF(no opt) across Re10K, ACID, and DL3DV. Combined gains reflect largely independent, additive contributions from each mechanism.}
    \label{fig:ablation}
\end{figure}

To assess the independent impact of the two core optimizations on the same hardware, we report ablation results relative to the SCARF (no optimization) baseline. This baseline retains the SCARF compute array, on-chip buffers, and control framework but disables FSDR and SAES, restoring the original serialized S2$\rightarrow$S3 execution flow, effectively serving as a vanilla accelerator. Fig.~\ref{fig:ablation} isolates the performance gains of FSDR and SAES. We evaluate three configurations: FSDR only, SAES only, and both enabled. FSDR alone yields a geometric mean speedup of 1.35$\times$ by contracting the depth-search space. SAES alone achieves a 1.26$\times$ speedup via tile-level path selection. Combined, they reach a 1.59$\times$ speedup indicating largely independent, additive benefits. The strongest synergy occurs on ACID (1.70$\times$) and the weakest on DL3DV (1.49$\times$), which matches the dataset-level trends reported above.

\paragraph{\textbf{FSDR Detailed Analysis}}

Table~\ref{tab:fsdr_detail} breaks down the mechanisms behind the 1.35$\times$ FSDR contribution. The \emph{Guided Rate} is the fraction of pixels with narrowed search spaces via cached depth anchors, which ranges from 58\% to 87\%. As the Guided Rate increases, \emph{Depth Evals Saved in S2} rises from 43\% to 65\%, while \emph{Feat. Buf. Red.} rises from 41\% to 62\%, indicating that the FSDR gain comes from reducing depth-candidate evaluation and LPDDR4X feature-buffer traffic. \emph{Top-1 Cov.} measures the fraction of guided local depth windows that still cover the top-1 candidate selected by full search. It remains between 99.83\% and 99.95\% across all model and dataset pairs. Thus, almost all full-search top-1 candidates remain inside the local windows, consistent with the stronger semantic consistency of DINOv2 features on roads, sky, and building planes. MVSplat has the lowest Guided Rate on DL3DV at 58\%, because lightweight features form less stable matches under complex occlusion and repeated textures. Its Top-1 Cov. still reaches 99.83\%, showing that complex scenes mainly reduce the triggering scope of guided search rather than shifting the triggered windows away from full-search results.

\begin{table}[t]
    \centering
    \caption{FSDR per-model statistics.}
    \label{tab:fsdr_detail}
    \resizebox{\columnwidth}{!}{
    \begin{tabular}{llcccc}
        \toprule
        \textbf{Model} & \textbf{Dataset} & \makecell{\textbf{Guided}\\\textbf{Rate}} & \makecell{\textbf{Top-1}\\\textbf{Cov.}} & \makecell{\textbf{Depth Evals}\\\textbf{Saved in S2 (M)}} & \makecell{\textbf{Feat. Buf.}\\\textbf{Red. (MB)}} \\
        \midrule
        \multirow{3}{*}{Tran}  & Re10K & 72.1\% & 99.91\% & 4.54\,(54.2\%) & 415\,(51.5\%) \\
                                    & ACID  & 76.0\% & 99.93\% & 4.78\,(57.4\%) & 438\,(54.4\%) \\
                                    & DL3DV & 65.0\% & 99.88\% & 4.09\,(48.9\%) & 374\,(46.5\%) \\
        \cmidrule(l){2-6}
        \multirow{3}{*}{MV}    & Re10K & 66.2\% & 99.90\% & 1.04\,(49.6\%) & 95\,(47.2\%)  \\
                                    & ACID  & 71.0\% & 99.87\% & 1.12\,(53.4\%) & 102\,(50.7\%) \\
                                    & DL3DV & 58.0\% & 99.83\% & 0.91\,(43.4\%) & 84\,(41.7\%)  \\
        \cmidrule(l){2-6}
        \multirow{3}{*}{Depth} & Re10K & 83.0\% & 99.89\% & 5.22\,(62.2\%) & 478\,(59.4\%) \\
                                    & ACID  & 87.0\% & 99.95\% & 5.47\,(65.2\%) & 501\,(62.3\%) \\
                                    & DL3DV & 78.0\% & 99.85\% & 4.91\,(58.5\%) & 449\,(55.8\%) \\
        \bottomrule
    \end{tabular}}
\end{table}

\paragraph{\textbf{SAES Detailed Analysis}}

Table~\ref{tab:saes_detail} similarly traces SAES's 1.26$\times$ contribution to tile-level path selection statistics. The \emph{L0 Rate} indicates the fraction of tiles entering the representative-generation path from 12.0\% to 46.7\%, while the \emph{L1 Rate} shows the fraction entering the lightweight path from 8.5\% to 25.1\%. Together, these rates determine the workload bypassing the Full path, leading to \emph{Gaussians Saved} from 16.1\% to 33.8\% and \emph{S2 Evals Saved} from 0.45M to 3.92M by skipping S2 search for L0 or L1 tiles. \emph{Low-Var. Agree.} measures the fraction of L0 or L1 tiles whose full per-pixel materialization already has low-variation Gaussian attributes. It remains between 99.23\% and 100.00\% across all combinations, showing that SAES mainly compresses Gaussian outputs that are already locally consistent under full materialization. On DL3DV, complex textures and depth boundaries keep more tiles on the Full path, reducing TranSplat's L0 Rate to 12.0\%. Its Low-Var. Agree. still reaches 99.23\%, indicating that complex regions mainly reduce early-path triggering rather than causing early paths to cover many unsuitable tiles.

\begin{table}[t]
    \centering
    \caption{SAES per-model statistics.}
    \label{tab:saes_detail}
    \resizebox{\columnwidth}{!}{
    \begin{tabular}{llccccc}
        \toprule
        \textbf{Model} & \textbf{Dataset} & \textbf{L0 Rate} & \textbf{L1 Rate} & \makecell{\textbf{Low-Var.}\\\textbf{Agree.}} & \makecell{\textbf{Gaussians}\\\textbf{Saved}} & \makecell{\textbf{S2 Evals}\\\textbf{Saved (M)}} \\
        \midrule
        \multirow{3}{*}{Tran}  & Re10K & 17.2\% & 14.8\% & 100.00\% & 20.2\% & 1.45 \\
                                    & ACID  & 21.3\% & 11.2\% & 99.86\% & 24.3\% & 1.79 \\
                                    & DL3DV & 12.0\% & 10.1\% & 99.23\% & 17.0\% & 1.01 \\
        \cmidrule(l){2-7}
        \multirow{3}{*}{MV}    & Re10K & 29.0\% & 12.5\% & 100.00\% & 24.0\% & 0.61 \\
                                    & ACID  & 36.0\% & 18.0\% & 99.77\% & 28.0\% & 0.76 \\
                                    & DL3DV & 21.3\% & 8.5\%  & 99.75\% & 16.1\% & 0.45 \\
        \cmidrule(l){2-7}
        \multirow{3}{*}{Depth} & Re10K & 38.7\% & 18.6\% & 99.59\% & 29.2\% & 3.25 \\
                                    & ACID  & 46.7\% & 25.1\% & 99.85\% & 33.8\% & 3.92 \\
                                    & DL3DV & 30.7\% & 13.2\% & 99.62\% & 24.8\% & 2.57 \\
        \bottomrule
    \end{tabular}}
\end{table}

\begin{figure}[t]
    \centering
    \includegraphics[width=\linewidth]{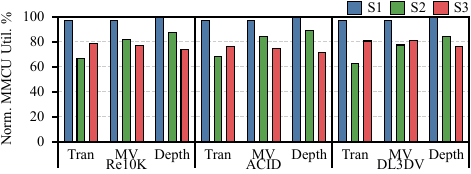}
    \caption{Normalized stage-wise MMCU active utilization.}
    \Description{Grouped bar chart showing normalized MMCU active utilization for S1, S2, and S3 across nine model and dataset pairs.}
    \label{fig:mmcu_utilization}
\end{figure}

\paragraph{\textbf{MMCU Utilization}}

Fig.~\ref{fig:mmcu_utilization} reports normalized MMCU active utilization for S1, S2, and S3. S1 reaches 97.9\% utilization, making backbone feature extraction a stable high-load stage. S2 averages 78.0\% because depth prediction mixes matrix refinement with cost-volume sampling, FSDR-guided search, and candidate reduction. TranSplat interleaves more sampling and reduction between MMCU kernels, MVSplat uses a lighter cost volume, and DepthSplat exposes denser mono-ViT refinement. S3 averages 76.7\% and follows SAES path selection. Representative and lightweight tiles reduce full Gaussian-head execution, so DepthSplat is lower on Re10K and ACID. S4 runs on the GGU Array and overlaps with the preceding three stages, requiring no additional MMCU operations. The shared MMCU therefore remains active across S1, S2, and S3 and preserves the area amortization benefit provided by the shared substrate.

\subsection{Sensitivity Study}

\paragraph{\textbf{FSDR Cache Size}}

\begin{figure}[t]
    \centering
    \includegraphics[width=\linewidth]{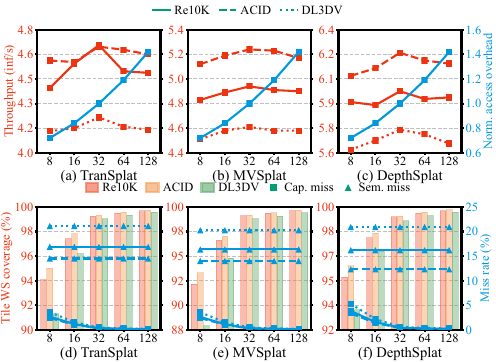}
    \caption{Sensitivity to FSDR cache size.}
    \Description{Per-model charts versus FSDR cache size from 8 to 128 entries across Re10K, ACID, and DL3DV. Panels (a)--(c) plot throughput and normalized access overhead. Panels (d)--(f) plot tile working-set coverage with capacity and semantic miss rates. Coverage saturates and capacity miss collapses by 32 entries, while semantic miss stays nearly constant.}
    \label{fig:sens_fsdr_cache}
\end{figure}

Fig.~\ref{fig:sens_fsdr_cache} relates FSDR cache size to throughput, access overhead, tile working-set coverage, and miss rate. Each cache entry stores one feature signature and its depth anchor, and the cache is reset at the beginning of each frame. Tile working-set coverage is the fraction of guided pixels whose reusable depth anchor is already resident when probed. From 8 to 128 entries, coverage rises while capacity miss falls. Both trends saturate by 32 entries, where geometric-mean throughput also peaks. Capacity miss drops by 1.48 percentage points from 16 to 32 entries, but only by 0.22 and 0.12 points over the next two doublings. At the same time, normalized access overhead rises from 1.00 at 32 entries to 1.19 and 1.42 at 64 and 128 entries, because a larger CAM raises per-lookup cost. Past 32 entries, the extra capacity gives little miss reduction, so access overhead becomes the main limiter and throughput declines slightly at 64 and 128 entries.  Fig.~\ref{fig:sens_fsdr_cache}(d)--(f) further shows that semantic miss stays near 26.07\% at all sizes, indicating that residual misses are set by scene content and feature similarity. Per-frame reset also has limited runtime impact. In our measurements, filling 32 distinct anchors takes 130 guided-pixel probes on average, only 0.198\% of a $256\times256$ view. Overall, 32 entries cover the main tile working set. Smaller caches increase capacity miss, while larger caches mainly increase lookup overhead, so we use 32 entries as the default.

\begin{figure}[t]
    \centering
    \includegraphics[width=\linewidth]{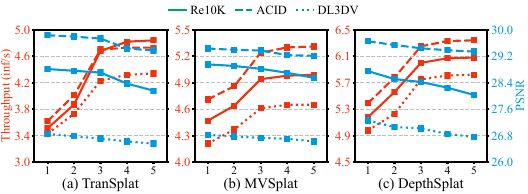}
    \caption{Sensitivity to FSDR Hamming threshold $\tau_h$.}
    \Description{Dual-axis line charts of throughput and quality versus $\tau_h$ (1--5) across all nine model--dataset combinations. Throughput saturates beyond $\tau_h=3$ while quality degrades monotonically.}
    \label{fig:sens_fsdr_thresh}
\end{figure}

\paragraph{\textbf{FSDR Hamming Threshold}}

Fig.~\ref{fig:sens_fsdr_thresh} illustrates the effect of the FSDR Hamming threshold, \(\tau_h\), on system performance. \(\tau_h\) sets how similar two pixel features must be before FSDR uses a cached depth to narrow the search range. Since this decision is local, scene changes mainly alter the number of pixels that satisfy the match condition, and therefore the FSDR hit rate and speedup. The threshold choice depends instead on the marginal reliability of the extra matches admitted as Hamming distance grows. Larger \(\tau_h\) first adds stable near-feature matches, then weaker matches with smaller throughput gains and higher quality risk. The same trend appears across Re10K, ACID, and DL3DV, indicating that the transition from reliable reuse to weak matching does not shift strongly with scene distribution. Increasing \(\tau_h\) from 1 to 3 raises geometric-mean throughput by about 18.1\%, while increasing it to 5 adds only 1.5\%, with a 0.20\% gain from 4 to 5. Quality fluctuation remains below 1.15\% at \(\tau_h=3\), but grows to 3.47\% at \(\tau_h=5\). We therefore set \(\tau_h=3\) by default to capture reliable reuse and avoid weak matches.

\paragraph{\textbf{SAES Variance Thresholds}}

\begin{figure}[t]
    \centering
    \includegraphics[width=\linewidth]{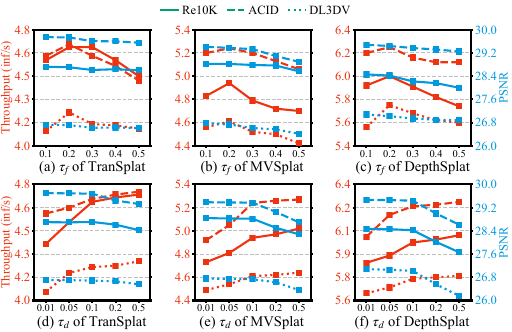}
    \caption{Sensitivity to SAES variance thresholds $\tau_f$ and $\tau_d$.}
    \Description{(a)--(c) line charts of throughput and quality versus $\tau_f$ (0.1--0.5). Throughput peaks unimodally at 0.20. (d)--(f) versus $\tau_d$ (0.01--0.5). Throughput increases monotonically and saturates beyond 0.1, while quality drops sharply at $\tau_d \ge 0.2$.}
    \label{fig:sens_saes_thresh}
\end{figure}

Fig.~\ref{fig:sens_saes_thresh}(a)--(c) evaluates the feature-variance threshold, \(\tau_f\), while panels (d)--(f) assess the depth-variance threshold, \(\tau_d\). Both thresholds are derived from probe statistics inside each tile. They measure local appearance consistency and local geometry consistency, tying the decision to tile-local image evidence rather than to a scene category or model type. Model and scene changes alter the prevalence of these local patterns, and therefore change the L0 and L1 rates and speedup. They do not significantly change the relation between variance and local approximability. The measurements follow this behavior. \(\tau_f\) shows a consistent unimodal trend across all nine combinations and peaks at 0.2. Further relaxing \(\tau_f\) routes more appearance-varying tiles to L0, increasing moment-matching overhead and quality fluctuation. For \(\tau_d\), throughput saturates beyond 0.1, while quality drops sharply at \(\tau_d \geq 0.2\). This indicates that the main locally smooth geometry has already been captured by L1 and that further relaxation mainly admits tiles with stronger depth variation. We therefore use \(\tau_f=0.20\) and \(\tau_d=0.1\) as defaults. They mark the boundary where local approximability changes from reliable to unreliable. 

\begin{figure}[t]
    \centering
    \includegraphics[width=\linewidth]{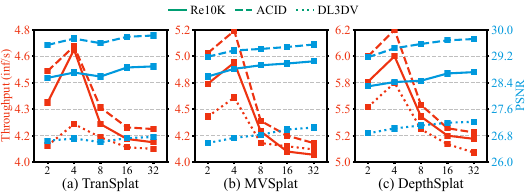}
    \caption{Sensitivity to SAES tile size.}
    \Description{Line charts of throughput and quality versus tile size (2$\times$2 to 32$\times$32) across all nine model--dataset combinations. $4\times4$ achieves the highest throughput with a clear peak-and-decline trend.}
    \label{fig:sens_tile}
\end{figure}

\paragraph{\textbf{Tile Size}}

Fig.~\ref{fig:sens_tile} illustrates the impact of tile size on the effectiveness of SAES. Throughput peaks at a $4\times4$ tile size across all nine model-dataset pairs. Both smaller and larger tiles reduce performance. The geometric mean throughput decreases by approximately 4.1\% for $2\times2$ tiles and by 9.6\%, 12.1\%, and 12.7\% for $8\times8$, $16\times16$, and $32\times32$ tiles, respectively, compared to $4\times4$. Smaller tiles improve local decision granularity but increase scheduling and path-switching overhead, offsetting computational gains. Larger tiles excessively average local texture and geometry, reducing L0/L1 early stopping triggers and causing more regions to revert to the Full path, thereby decreasing pruning efficiency. Quality generally improves with tile size. The $32\times32$ setting yields the highest quality, only 1.67\% better than $4\times4$, but incurs over a 12\% throughput loss. Balancing quality and performance, $4\times4$ is selected as the default tile size.

\subsection{Area and Power Analysis}

\paragraph{\textbf{Area Analysis}}

As shown in Table~\ref{tab:area_power}, SCARF occupies 27.8 mm$^2$ in 28 nm, with the area distribution directly reflecting the workload hierarchy. The MMCU dominates at 50.4\% (14.0 mm$^2$) because it serves as the shared compute substrate across S1 through S3. Concentrating compute into one $48\times48$ array avoids the duplication of stage-specific designs. The GGU Array accounts for 17.3\% (4.80 mm$^2$), sized to run its 32 geometry-conversion PEs concurrently with the MVU. The FSDR subsystem occupies only 0.08 mm$^2$ (0.3\%), including only its 32-entry CAM, confirming that semantic indexing imposes negligible silicon cost relative to the 41\% to 62\% DRAM traffic reduction it enables. On-chip buffers (1.26 mm$^2$, 4.5\%) sustain inter-stage data reuse, while the LPDDR4X PHY (3.50\,mm$^2$, 12.6\%) reflects the unavoidable fixed analog cost of high-speed I/O.

\paragraph{\textbf{Power Analysis}}

In Table~\ref{tab:area_power}, SCARF consumes 2.1 W at 1 GHz and 1.0 V, enabling fanless operation. Dynamic power mirrors the area hierarchy: the MMCU dominates at 52.2\% (1,016 mW) as the primary active-compute fabric, while I/O and PHY account for 12.5\% (244.0 mW) as the fixed cost of off-chip bandwidth. Despite consuming only 15 mW ($<$1\%), the FSDR subsystem delivers outsized savings. By cutting DRAM traffic by 41\% to 62\%, it suppresses the I/O switching energy that would otherwise grow with every redundant depth-search cycle. SAES compounds this effect through tile-level path pruning, reducing S3 dynamic power proportionally to the Gaussian reduction in Table~\ref{tab:saes_detail}.

\begin{table}[t]
    \centering
    \caption{SCARF area and power breakdown in 28\,nm.}
    \label{tab:area_power}
    \resizebox{\columnwidth}{!}{
    \begin{tabular}{lccc}
        \toprule
        \textbf{Component} & \textbf{Area\,(mm$^2$)} & \textbf{Dyn.\,(mW)} & \textbf{Stat.\,(mW)} \\
        \midrule
        \multicolumn{4}{l}{\textit{MVU\enspace(15.35\,mm$^2$,\ 1{,}133.6\,/\,82.3\,mW)}} \\
        \quad MMCU                        & 14.0\,(50.4\%) & 1,016\,(52.2\%)  & 73.8\,(50.9\%) \\
        \quad VectorALU                   & 0.50\,(1.8\%)  & 46.7\,(2.4\%)    & 2.1\,(1.4\%)   \\
        \quad BilinearUnit                & 0.82\,(3.0\%)  & 61.9\,(3.2\%)    & 4.8\,(3.3\%)   \\
        \quad NormUnit                    & 0.02\,(0.1\%)  & 5.8\,(0.30\%)    & 0.9\,(0.62\%)  \\
        \quad ActivationUnit              & 0.01\,(0.04\%) & 3.2\,(0.16\%)    & 0.7\,(0.48\%)  \\
        \midrule
        \multicolumn{4}{l}{\textit{GGU Array ($\times$32 PEs)\enspace(4.80\,mm$^2$,\ 314.6\,/\,25.6\,mW)}} \\
        \quad PositionCalc                & 0.65\,(2.3\%)  & 42.1\,(2.2\%)    & 3.4\,(2.3\%)   \\
        \quad CovBuilder                  & 1.35\,(4.9\%)  & 89.3\,(4.6\%)    & 7.2\,(5.0\%)   \\
        \quad SH\_OPGenerator             & 2.80\,(10.1\%) & 183.2\,(9.4\%)   & 15.0\,(10.3\%) \\
        \midrule
        \multicolumn{4}{l}{\textit{FSDR Subsystem\enspace(0.08\,mm$^2$,\ 15.3\,/\,1.2\,mW)}} \\
        \quad LSHHashUnit                 & 0.03\,(0.1\%)  & 6.1\,(0.31\%)    & 0.5\,(0.34\%)  \\
        \quad CAM Array (32-entry)        & 0.04\,(0.1\%)  & 7.8\,(0.40\%)    & 0.5\,(0.34\%)  \\
        \quad FSDR Controller             & 0.01\,(0.04\%) & 1.4\,(0.07\%)    & 0.2\,(0.14\%)  \\
        \midrule
        \multicolumn{4}{l}{\textit{On-chip Buffers\enspace(1.26\,mm$^2$,\ 153.3\,/\,18.7\,mW)}} \\
        \quad Weight Buffer (128\,KB)     & 0.38\,(1.4\%)  & 41.2\,(2.1\%)    & 6.3\,(4.3\%)   \\
        \quad Feature Buffer (256\,KB)    & 0.70\,(2.5\%)  & 89.4\,(4.6\%)    & 10.1\,(7.0\%)  \\
        \quad Tile Buffer (64\,KB)        & 0.18\,(0.6\%)  & 22.7\,(1.2\%)    & 2.3\,(1.6\%)   \\
        \midrule
        \multicolumn{4}{l}{\textit{Control \& Clock\enspace(0.47\,mm$^2$,\ 86.2\,/\,5.2\,mW)}} \\
        \quad Control + Interconnect      & 0.32\,(1.2\%)  & 27.8\,(1.4\%)    & 1.4\,(1.0\%)   \\
        \quad PLL + Clock tree            & 0.15\,(0.5\%)  & 58.4\,(3.0\%)    & 3.8\,(2.6\%)   \\
        \midrule
        \multicolumn{4}{l}{\textit{I/O \& PHY\enspace(3.50\,mm$^2$,\ 244.0\,/\,12.0\,mW)}} \\
        \quad I/O + LPDDR4X PHY            & 3.50\,(12.6\%) & 244.0\,(12.5\%)  & 12.0\,(8.3\%)  \\
        \midrule
        Routing / filler (85\% util.)     & 2.32\,(8.3\%)  & ---              & ---            \\
        \midrule
        \textbf{Total die}                & \textbf{27.8\,mm$^2$} & \multicolumn{2}{c}{\textbf{2{,}092\,mW}} \\
        \bottomrule
    \end{tabular}}
\end{table}

\section{Related Work}

\paragraph{\textbf{Feed-forward 3D reconstruction and 3DGS rendering accelerators}}

Recent accelerators target feed-forward 3D reconstruction and 3DGS rendering workloads. VersaQ-3D~\cite{zhang2026versaq3d} accelerates VGGT-based generalizable 3D reconstruction through versatile quantization and a reconfigurable datapath for transformer inference. GCC~\cite{pei2025gcc} optimizes the coupling between preprocessing and rendering with cross-stage conditional processing and Gaussian-wise rendering, reducing unnecessary preprocessing and repeated Gaussian loading. Neo~\cite{oh2026neo} targets real-time on-device 3DGS rendering and exploits temporal redundancy in depth ordering across consecutive frames through reuse-and-update sorting. These works optimize either transformer-based reconstruction or rendering from an existing Gaussian set. SCARF targets depth-guided, cost-volume-based G-3DGS encoder inference, where cost-volume depth search and Gaussian materialization dominate encoder construction. By reducing this encoder-side work before rendering, SCARF sits upstream of rasterization and complements GCC and Neo.

\paragraph{\textbf{3DGS memory scaling and training systems}}

A separate line of systems work addresses memory capacity limits in large-scale 3DGS reconstruction and training. CLM~\cite{zhao2026clm} offloads Gaussian parameters to CPU memory and loads the active working set into GPU memory when needed, using pipelining and access-pattern-aware communication reduction to support very large scenes on a single consumer GPU. GS-Scale~\cite{lee2026gsscale} similarly stores Gaussian parameters, gradients, and optimizer states in host memory, then reduces training overhead with selective offloading, parameter forwarding, and deferred optimizer updates. These systems expand the scale of per-scene 3DGS optimization and rendering, but they do not change the computation inside feed-forward G-3DGS encoders. SCARF targets this different stage by reducing cost-volume depth search and redundant Gaussian materialization before the final Gaussian set is emitted.

\section{Conclusion}

SCARF targets inference in depth-guided G-3DGS encoders with cost-volume depth prediction. It exploits semantic locality through Feature Similarity Depth Reuse, which narrows depth candidates by feature-space affinity, and local geometric redundancy through Scene-Adaptive Early Sparsification, which activates the minimal materialization path before full primitive instantiation. Across three depth-guided, cost-volume-based models and three datasets, SCARF delivers a 2.94$\times$ geometric-mean end-to-end speedup, 2.23 inferences per joule, and a maximum PSNR drop under 0.072\%. Implemented in 28 nm, SCARF occupies 27.8 mm$^2$ and consumes 2.1 W, demonstrating practical potential for edge novel-view synthesis.

For downstream integration, SCARF serves as the Gaussian producer before a rasterizer. S4 writes primitives to a buffer that an existing 3DGS rasterizer reads through its input interface. For an end-to-end G-3DGS pipeline, future work can study cross-stage scheduling, double buffering, and on-chip reuse at this boundary.

\bibliographystyle{IEEEtran}
\bibliography{sample-base}

@inproceedings{ajayi2019openroad,
  author    = {Ajayi, Tutu and Chhabria, Vidya A. and Foga\c{c}a, Mateus and Hashemi, Soheil and Hosny, Abdelrahman and Kahng, Andrew B. and Kim, Minsoo and Lee, Jeongsup and Mallappa, Uday and Neseem, Marina and Pradipta, Geraldo and Reda, Sherief and Saligane, Mehdi and Sapatnekar, Sachin S. and Sechen, Carl and Shalan, Mohamed and Swartz, William and Wang, Lutong and Wang, Zhehong and Woo, Mingyu and Xu, Bangqi},
  title     = {Toward an Open-Source Digital Flow: First Learnings from the {OpenROAD} Project},
  year      = {2019},
  isbn      = {9781450367257},
  publisher = {Association for Computing Machinery},
  address   = {New York, NY, USA},
  url       = {https://doi.org/10.1145/3316781.3326334},
  doi       = {10.1145/3316781.3326334},
  booktitle = {Proceedings of the 56th Annual Design Automation Conference 2019},
  articleno = {76},
  numpages  = {4},
  location  = {Las Vegas, NV, USA},
  series    = {DAC '19}
}

@inproceedings{caron2021dino,
  author    = {Caron, Mathilde and Touvron, Hugo and Misra, Ishan and Jegou, Hervé and Mairal, Julien and Bojanowski, Piotr and Joulin, Armand},
  booktitle = {2021 IEEE/CVF International Conference on Computer Vision (ICCV)},
  title     = {Emerging Properties in Self-Supervised Vision Transformers},
  year      = {2021},
  volume    = {},
  number    = {},
  pages     = {9630-9640},
  doi       = {10.1109/ICCV48922.2021.00951}
}

@inproceedings{charatan23pixelsplat,
  author    = {David Charatan and
               Sizhe Lester Li and
               Andrea Tagliasacchi and
               Vincent Sitzmann},
  title     = {{pixelSplat}: {3D} Gaussian Splats from Image Pairs for Scalable Generalizable
               {3D} Reconstruction},
  booktitle = {{IEEE/CVF} Conference on Computer Vision and Pattern Recognition,
               {CVPR} 2024, Seattle, WA, USA, June 16-22, 2024},
  pages     = {19457--19467},
  publisher = {{IEEE}},
  year      = {2024},
  url       = {https://doi.org/10.1109/CVPR52733.2024.01840},
  doi       = {10.1109/CVPR52733.2024.01840},
  bibsource = {dblp computer science bibliography, https://dblp.org}
}

@inproceedings{charikar2002similarity-estimation,
  author    = {Charikar, Moses S.},
  title     = {Similarity estimation techniques from rounding algorithms},
  year      = {2002},
  isbn      = {1581134959},
  publisher = {Association for Computing Machinery},
  address   = {New York, NY, USA},
  url       = {https://doi.org/10.1145/509907.509965},
  doi       = {10.1145/509907.509965},
  booktitle = {Proceedings of the Thirty-Fourth Annual ACM Symposium on Theory of Computing},
  pages     = {380–388},
  numpages  = {9},
  location  = {Montreal, Quebec, Canada},
  series    = {STOC '02}
}

@inproceedings{chen2021mvsnerf,
  author    = {Chen, Anpei and Xu, Zexiang and Zhao, Fuqiang and Zhang, Xiaoshuai and Xiang, Fanbo and Yu, Jingyi and Su, Hao},
  booktitle = {2021 IEEE/CVF International Conference on Computer Vision (ICCV)},
  title     = {{MVSNeRF}: Fast Generalizable Radiance Field Reconstruction from Multi-View Stereo},
  year      = {2021},
  volume    = {},
  number    = {},
  pages     = {14104-14113},
  doi       = {10.1109/ICCV48922.2021.01386}
}

@inproceedings{chen2024mvsplat,
  author    = {Chen, Yuedong
               and Xu, Haofei
               and Zheng, Chuanxia
               and Zhuang, Bohan
               and Pollefeys, Marc
               and Geiger, Andreas
               and Cham, Tat-Jen
               and Cai, Jianfei},
  editor    = {Leonardis, Ale{\v{s}}
               and Ricci, Elisa
               and Roth, Stefan
               and Russakovsky, Olga
               and Sattler, Torsten
               and Varol, G{\"u}l},
  title     = {{MVSplat}: Efficient {3D} Gaussian Splatting from Sparse Multi-view Images},
  booktitle = {Computer Vision -- ECCV 2024},
  year      = {2025},
  publisher = {Springer Nature Switzerland},
  address   = {Cham},
  pages     = {370--386},
  isbn      = {978-3-031-72664-4}
}

@article{chetlur2014cudnn,
  author     = {Sharan Chetlur and
                Cliff Woolley and
                Philippe Vandermersch and
                Jonathan Cohen and
                John Tran and
                Bryan Catanzaro and
                Evan Shelhamer},
  title      = {{cuDNN}: Efficient Primitives for Deep Learning},
  journal    = {CoRR},
  volume     = {abs/1410.0759},
  year       = {2014},
  url        = {http://arxiv.org/abs/1410.0759},
  eprinttype = {arXiv},
  eprint     = {1410.0759},
  bibsource  = {dblp computer science bibliography, https://dblp.org}
}

@article{comaniciu2002meanshift,
  author     = {Comaniciu, Dorin and Meer, Peter},
  title      = {Mean Shift: A Robust Approach Toward Feature Space Analysis},
  year       = {2002},
  issue_date = {May 2002},
  publisher  = {IEEE Computer Society},
  address    = {USA},
  volume     = {24},
  number     = {5},
  issn       = {0162-8828},
  url        = {https://doi.org/10.1109/34.1000236},
  doi        = {10.1109/34.1000236},
  journal    = {IEEE Trans. Pattern Anal. Mach. Intell.},
  month      = may,
  pages      = {603–619},
  numpages   = {17}
}

@inproceedings{ding2022transmvsnet,
  author    = {Ding, Yikang and Yuan, Wentao and Zhu, Qingtian and Zhang, Haotian and Liu, Xiangyue and Wang, Yuanjiang and Liu, Xiao},
  booktitle = {2022 IEEE/CVF Conference on Computer Vision and Pattern Recognition (CVPR)},
  title     = {{TransMVSNet}: Global Context-aware Multi-view Stereo Network with Transformers},
  year      = {2022},
  volume    = {},
  number    = {},
  pages     = {8575-8584},
  doi       = {10.1109/CVPR52688.2022.00839}
}

@inproceedings{dosovitskiy2020vit,
  author    = {Alexey Dosovitskiy and
               Lucas Beyer and
               Alexander Kolesnikov and
               Dirk Weissenborn and
               Xiaohua Zhai and
               Thomas Unterthiner and
               Mostafa Dehghani and
               Matthias Minderer and
               Georg Heigold and
               Sylvain Gelly and
               Jakob Uszkoreit and
               Neil Houlsby},
  title     = {An Image is Worth 16x16 Words: Transformers for Image Recognition
               at Scale},
  booktitle = {9th International Conference on Learning Representations, {ICLR} 2021,
               Virtual Event, Austria, May 3-7, 2021},
  publisher = {OpenReview.net},
  year      = {2021},
  url       = {https://openreview.net/forum?id=YicbFdNTTy},
  bibsource = {dblp computer science bibliography, https://dblp.org}
}

@inproceedings{fan2024lightgaussian,
  author    = {Fan, Zhiwen and Wang, Kevin and Wen, Kairun and Zhu, Zehao and Xu, Dejia and Wang, Zhangyang},
  booktitle = {Advances in Neural Information Processing Systems},
  doi       = {10.52202/079017-4447},
  editor    = {A. Globerson and L. Mackey and D. Belgrave and A. Fan and U. Paquet and J. Tomczak and C. Zhang},
  pages     = {140138--140158},
  publisher = {Curran Associates, Inc.},
  title     = {{LightGaussian}: Unbounded {3D} Gaussian Compression with 15x Reduction and 200+ {FPS}},
  url       = {https://proceedings.neurips.cc/paper_files/paper/2024/file/fd881d3b625437354d4421818f81058f-Paper-Conference.pdf},
  volume    = {37},
  year      = {2024}
}

@inproceedings{he2016deep-residual-learning,
  author    = {Kaiming He and
               Xiangyu Zhang and
               Shaoqing Ren and
               Jian Sun},
  title     = {Deep Residual Learning for Image Recognition},
  booktitle = {2016 {IEEE} Conference on Computer Vision and Pattern Recognition,
               {CVPR} 2016, Las Vegas, NV, USA, June 27-30, 2016},
  pages     = {770--778},
  publisher = {{IEEE} Computer Society},
  year      = {2016},
  url       = {https://doi.org/10.1109/CVPR.2016.90},
  doi       = {10.1109/CVPR.2016.90},
  bibsource = {dblp computer science bibliography, https://dblp.org}
}

@inproceedings{huang2024photoslam,
  author    = {Huajian Huang and
               Longwei Li and
               Hui Cheng and
               Sai{-}Kit Yeung},
  title     = {{Photo-SLAM}: Real-Time Simultaneous Localization and Photorealistic
               Mapping for Monocular, Stereo, and {RGB-D} Cameras},
  booktitle = {{IEEE/CVF} Conference on Computer Vision and Pattern Recognition,
               {CVPR} 2024, Seattle, WA, USA, June 16-22, 2024},
  pages     = {21584--21593},
  publisher = {{IEEE}},
  year      = {2024},
  url       = {https://doi.org/10.1109/CVPR52733.2024.02039},
  doi       = {10.1109/CVPR52733.2024.02039},
  bibsource = {dblp computer science bibliography, https://dblp.org}
}

@inproceedings{kendall2017endtoendlo,
  author    = {Kendall, Alex and Martirosyan, Hayk and Dasgupta, Saumitro and Henry, Peter and Kennedy, Ryan and Bachrach, Abraham and Bry, Adam},
  booktitle = {2017 IEEE International Conference on Computer Vision (ICCV)},
  title     = {End-to-End Learning of Geometry and Context for Deep Stereo Regression},
  year      = {2017},
  volume    = {},
  number    = {},
  pages     = {66-75},
  doi       = {10.1109/ICCV.2017.17}
}

@article{kerbl20233dgs,
  author     = {Kerbl, Bernhard and Kopanas, Georgios and Leimkuehler, Thomas and Drettakis, George},
  title      = {{3D} Gaussian Splatting for Real-Time Radiance Field Rendering},
  year       = {2023},
  issue_date = {August 2023},
  publisher  = {Association for Computing Machinery},
  address    = {New York, NY, USA},
  volume     = {42},
  number     = {4},
  issn       = {0730-0301},
  url        = {https://doi.org/10.1145/3592433},
  doi        = {10.1145/3592433},
  journal    = {ACM Trans. Graph.},
  month      = jul,
  articleno  = {139},
  numpages   = {14}
}

@inproceedings{lee2024gscore,
  author    = {Lee, Junseo and Lee, Seokwon and Lee, Jungi and Park, Junyong and Sim, Jaewoong},
  title     = {{GSCore}: Efficient Radiance Field Rendering via Architectural Support for {3D} Gaussian Splatting},
  year      = {2024},
  isbn      = {9798400703867},
  publisher = {Association for Computing Machinery},
  address   = {New York, NY, USA},
  url       = {https://doi.org/10.1145/3620666.3651385},
  doi       = {10.1145/3620666.3651385},
  booktitle = {Proceedings of the 29th ACM International Conference on Architectural Support for Programming Languages and Operating Systems, Volume 3},
  pages     = {497–511},
  numpages  = {15},
  location  = {La Jolla, CA, USA},
  series    = {ASPLOS '24}
}

@inproceedings{oh2026neo,
  author    = {Oh, Changhun and Oh, Seongryong and Hwang, Jinwoo and Kim, Yoonsung and Sharma, Hardik and Park, Jongse},
  title     = {Neo: Real-Time On-Device {3D} Gaussian Splatting with Reuse-and-Update Sorting Acceleration},
  year      = {2026},
  publisher = {Association for Computing Machinery},
  address   = {New York, NY, USA},
  url       = {https://doi.org/10.1145/3779212.3790192},
  doi       = {10.1145/3779212.3790192},
  booktitle = {Proceedings of the 31st ACM International Conference on Architectural Support for Programming Languages and Operating Systems, Volume 2},
  pages     = {1268--1284},
  numpages  = {17},
  series    = {ASPLOS 2026}
}

@misc{zhang2026versaq3d,
  author        = {Zhang, Yipu and Cheng, Jintao and Liu, Xingyu and Li, Zeyu and Li, Carol Jingyi and Wu, Jin and Jiang, Lin and Xie, Yuan and Xu, Jiang and Zhang, Wei},
  title         = {{VersaQ-3D}: A Reconfigurable Accelerator Enabling Feed-Forward and Generalizable {3D} Reconstruction via Versatile Quantization},
  year          = {2026},
  eprint        = {2601.20317},
  archiveprefix = {arXiv},
  primaryclass  = {cs.AR},
  url           = {https://arxiv.org/abs/2601.20317}
}

@inproceedings{li2023instant-3d,
  author    = {Li, Sixu and Li, Chaojian and Zhu, Wenbo and Yu, Boyang (Tony) and Zhao, Yang (Katie) and Wan, Cheng and You, Haoran and Shi, Huihong and Lin, Yingyan (Celine)},
  title     = {{Instant-3D}: Instant Neural Radiance Field Training Towards On-Device {AR/VR} {3D} Reconstruction},
  year      = {2023},
  isbn      = {9798400700958},
  publisher = {Association for Computing Machinery},
  address   = {New York, NY, USA},
  url       = {https://doi.org/10.1145/3579371.3589115},
  doi       = {10.1145/3579371.3589115},
  booktitle = {Proceedings of the 50th Annual International Symposium on Computer Architecture},
  articleno = {6},
  numpages  = {13},
  location  = {Orlando, FL, USA},
  series    = {ISCA '23}
}

@inproceedings{ling2024dl3dv,
  author    = {Lu Ling and
               Yichen Sheng and
               Zhi Tu and
               Wentian Zhao and
               Cheng Xin and
               Kun Wan and
               Lantao Yu and
               Qianyu Guo and
               Zixun Yu and
               Yawen Lu and
               Xuanmao Li and
               Xingpeng Sun and
               Rohan Ashok and
               Aniruddha Mukherjee and
               Hao Kang and
               Xiangrui Kong and
               Gang Hua and
               Tianyi Zhang and
               Bedrich Benes and
               Aniket Bera},
  title     = {{DL3DV-10K:} {A} Large-Scale Scene Dataset for Deep Learning-based
               3D Vision},
  booktitle = {{IEEE/CVF} Conference on Computer Vision and Pattern Recognition,
               {CVPR} 2024, Seattle, WA, USA, June 16-22, 2024},
  pages     = {22160--22169},
  publisher = {{IEEE}},
  year      = {2024},
  url       = {https://doi.org/10.1109/CVPR52733.2024.02092},
  doi       = {10.1109/CVPR52733.2024.02092},
  bibsource = {dblp computer science bibliography, https://dblp.org}
}

@inproceedings{liu2021infinite-nature,
  author    = {Liu, Andrew and Makadia, Ameesh and Tucker, Richard and Snavely, Noah and Jampani, Varun and Kanazawa, Angjoo},
  booktitle = {2021 IEEE/CVF International Conference on Computer Vision (ICCV)},
  title     = {Infinite Nature: Perpetual View Generation of Natural Scenes from a Single Image},
  year      = {2021},
  volume    = {},
  number    = {},
  pages     = {14438-14447},
  doi       = {10.1109/ICCV48922.2021.01419}
}

@inproceedings{lu2024scaffoldgs,
  author    = {Lu, Tao and Yu, Mulin and Xu, Linning and Xiangli, Yuanbo and Wang, Limin and Lin, Dahua and Dai, Bo},
  booktitle = {2024 IEEE/CVF Conference on Computer Vision and Pattern Recognition (CVPR)},
  title     = {{Scaffold-GS}: Structured {3D} Gaussians for View-Adaptive Rendering},
  year      = {2024},
  volume    = {},
  number    = {},
  pages     = {20654-20664},
  doi       = {10.1109/CVPR52733.2024.01952}
}

@article{luo2023ramulator2,
  author     = {Luo, Haocong and Tu\u{g}rul, Yahya Can and Bostanc\i{}, F. Nisa and Olgun, Ataberk and Ya\u{g}l\i{}k\c{c}\i{}, A. Giray and Mutlu, Onur},
  title      = {Ramulator 2.0: A Modern, Modular, and Extensible {DRAM} Simulator},
  year       = {2024},
  issue_date = {Jan.-June 2024},
  publisher  = {IEEE Computer Society},
  address    = {USA},
  volume     = {23},
  number     = {1},
  issn       = {1556-6056},
  url        = {https://doi.org/10.1109/LCA.2023.3333759},
  doi        = {10.1109/LCA.2023.3333759},
  journal    = {IEEE Comput. Archit. Lett.},
  month      = jan,
  pages      = {112–116},
  numpages   = {5}
}

@inproceedings{matsuki2023gaussian-splatting-slam,
  author    = {Matsuki, Hidenobu and Murai, Riku and Kelly, Paul H. J. and Davison, Andrew J.},
  booktitle = {2024 IEEE/CVF Conference on Computer Vision and Pattern Recognition (CVPR)},
  title     = {Gaussian Splatting {SLAM}},
  year      = {2024},
  volume    = {},
  number    = {},
  pages     = {18039-18048},
  doi       = {10.1109/CVPR52733.2024.01708}
}

@article{muller2022instant-nngp,
  author     = {M\"{u}ller, Thomas and Evans, Alex and Schied, Christoph and Keller, Alexander},
  title      = {Instant neural graphics primitives with a multiresolution hash encoding},
  year       = {2022},
  issue_date = {July 2022},
  publisher  = {Association for Computing Machinery},
  address    = {New York, NY, USA},
  volume     = {41},
  number     = {4},
  issn       = {0730-0301},
  url        = {https://doi.org/10.1145/3528223.3530127},
  doi        = {10.1145/3528223.3530127},
  journal    = {ACM Trans. Graph.},
  month      = jul,
  articleno  = {102},
  numpages   = {15}
}

@inproceedings{noh2025flexnerfer,
  author    = {Noh, Seock-Hwan and Shin, Banseok and Choi, Jeik and Lee, Seungpyo and Kung, Jaeha and Kim, Yeseong},
  title     = {{FlexNeRFer}: A Multi-Dataflow, Adaptive Sparsity-Aware Accelerator for On-Device {NeRF} Rendering},
  year      = {2025},
  isbn      = {9798400712616},
  publisher = {Association for Computing Machinery},
  address   = {New York, NY, USA},
  url       = {https://doi.org/10.1145/3695053.3731107},
  doi       = {10.1145/3695053.3731107},
  booktitle = {Proceedings of the 52nd Annual International Symposium on Computer Architecture},
  pages     = {1894–1909},
  numpages  = {16},
  location  = {
               },
  series    = {ISCA '25}
}

@article{oquab2024dinov2,
  author    = {Maxime Oquab and
               Timoth{\'{e}}e Darcet and
               Th{\'{e}}o Moutakanni and
               Huy V. Vo and
               Marc Szafraniec and
               Vasil Khalidov and
               Pierre Fernandez and
               Daniel Haziza and
               Francisco Massa and
               Alaaeldin El{-}Nouby and
               Mido Assran and
               Nicolas Ballas and
               Wojciech Galuba and
               Russell Howes and
               Po{-}Yao Huang and
               Shang{-}Wen Li and
               Ishan Misra and
               Michael Rabbat and
               Vasu Sharma and
               Gabriel Synnaeve and
               Hu Xu and
               Herv{\'{e}} J{\'{e}}gou and
               Julien Mairal and
               Patrick Labatut and
               Armand Joulin and
               Piotr Bojanowski},
  title     = {{DINOv2}: Learning Robust Visual Features without Supervision},
  journal   = {Trans. Mach. Learn. Res.},
  volume    = {2024},
  year      = {2024},
  url       = {https://openreview.net/forum?id=a68SUt6zFt},
  bibsource = {dblp computer science bibliography, https://dblp.org}
}

@inproceedings{paszke2019pytorch,
  author    = {Adam Paszke and
               Sam Gross and
               Francisco Massa and
               Adam Lerer and
               James Bradbury and
               Gregory Chanan and
               Trevor Killeen and
               Zeming Lin and
               Natalia Gimelshein and
               Luca Antiga and
               Alban Desmaison and
               Andreas K{\"{o}}pf and
               Edward Z. Yang and
               Zachary DeVito and
               Martin Raison and
               Alykhan Tejani and
               Sasank Chilamkurthy and
               Benoit Steiner and
               Lu Fang and
               Junjie Bai and
               Soumith Chintala},
  editor    = {Hanna M. Wallach and
               Hugo Larochelle and
               Alina Beygelzimer and
               Florence d'Alch{\'{e}}{-}Buc and
               Emily B. Fox and
               Roman Garnett},
  title     = {{PyTorch}: An Imperative Style, High-Performance Deep Learning Library},
  booktitle = {Advances in Neural Information Processing Systems 32: Annual Conference
               on Neural Information Processing Systems 2019, NeurIPS 2019, December
               8-14, 2019, Vancouver, BC, Canada},
  pages     = {8024--8035},
  year      = {2019},
  url       = {https://proceedings.neurips.cc/paper/2019/hash/bdbca288fee7f92f2bfa9f7012727740-Abstract.html},
  bibsource = {dblp computer science bibliography, https://dblp.org}
}

@inbook{pei2025gcc,
  author    = {Pei, Minnan and Li, Gang and Si, Junwen and Zhu, Zeyu and Mo, Zitao and Wang, Peisong and Song, Zhuoran and Liang, Xiaoyao and Cheng, Jian},
  title     = {{GCC}: A {3DGS} Inference Architecture with Gaussian-Wise and Cross-Stage Conditional Processing},
  year      = {2025},
  isbn      = {9798400715730},
  publisher = {Association for Computing Machinery},
  address   = {New York, NY, USA},
  url       = {https://doi.org/10.1145/3725843.3756072},
  booktitle = {Proceedings of the 58th IEEE/ACM International Symposium on Microarchitecture},
  pages     = {1824–1837},
  numpages  = {14}
}

@inproceedings{zhao2026clm,
  author    = {Zhao, Hexu and Min, Xiwen and Liu, Xiaoteng and Gong, Moonjun and Li, Yiming and Li, Ang and Xie, Saining and Li, Jinyang and Panda, Aurojit},
  title     = {{CLM}: Removing the {GPU} Memory Barrier for {3D} Gaussian Splatting},
  year      = {2026},
  publisher = {Association for Computing Machinery},
  address   = {New York, NY, USA},
  url       = {https://doi.org/10.1145/3779212.3790140},
  doi       = {10.1145/3779212.3790140},
  booktitle = {Proceedings of the 31st ACM International Conference on Architectural Support for Programming Languages and Operating Systems, Volume 2},
  pages     = {377--393},
  numpages  = {17},
  series    = {ASPLOS 2026}
}

@inproceedings{lee2026gsscale,
  author    = {Lee, Donghyun and Jeong, Dawoon and Lee, Jae W. and Yoon, Hongil},
  title     = {{GS-Scale}: Unlocking Large-Scale {3D} Gaussian Splatting Training via Host Offloading},
  year      = {2026},
  publisher = {Association for Computing Machinery},
  address   = {New York, NY, USA},
  url       = {https://doi.org/10.1145/3779212.3790167},
  doi       = {10.1145/3779212.3790167},
  booktitle = {Proceedings of the 31st ACM International Conference on Architectural Support for Programming Languages and Operating Systems, Volume 2},
  pages     = {860--875},
  numpages  = {16},
  series    = {ASPLOS 2026}
}

@inproceedings{ronneberger2015unet,
  author    = {Ronneberger, Olaf
               and Fischer, Philipp
               and Brox, Thomas},
  editor    = {Navab, Nassir
               and Hornegger, Joachim
               and Wells, William M.
               and Frangi, Alejandro F.},
  title     = {{U-Net}: Convolutional Networks for Biomedical Image Segmentation},
  booktitle = {Medical Image Computing and Computer-Assisted Intervention -- MICCAI 2015},
  year      = {2015},
  publisher = {Springer International Publishing},
  address   = {Cham},
  pages     = {234--241},
  isbn      = {978-3-319-24574-4}
}

@inproceedings{schonberger2016structure-from-motion-revisited,
  author    = {Schönberger, Johannes L. and Frahm, Jan-Michael},
  booktitle = {2016 IEEE Conference on Computer Vision and Pattern Recognition (CVPR)},
  title     = {{Structure-from-Motion} Revisited},
  year      = {2016},
  volume    = {},
  number    = {},
  pages     = {4104-4113},
  doi       = {10.1109/CVPR.2016.445}
}

@inproceedings{steiner2025drampower5,
  author    = {Steiner, Lukas and Psota, Thomas and M\"{o}rz, Marco and Christ, Derek and Jung, Matthias and Wehn, Norbert},
  title     = {{DRAMPower} 5: An Open-Source Power Simulator for Current Generation {DRAM} Standards},
  year      = {2025},
  isbn      = {9798400714719},
  publisher = {Association for Computing Machinery},
  address   = {New York, NY, USA},
  url       = {https://doi.org/10.1145/3721848.3721850},
  doi       = {10.1145/3721848.3721850},
  booktitle = {Proceedings of the Rapid Simulation and Performance Evaluation for Design Workshop},
  pages     = {8–16},
  numpages  = {9},
  location  = {
               },
  series    = {RAPIDO '25}
}

@inproceedings{sun2021loftr,
  author    = {Jiaming Sun and
               Zehong Shen and
               Yuang Wang and
               Hujun Bao and
               Xiaowei Zhou},
  title     = {{LoFTR}: Detector-Free Local Feature Matching With Transformers},
  booktitle = {{IEEE} Conference on Computer Vision and Pattern Recognition, {CVPR}
               2021, virtual, June 19-25, 2021},
  pages     = {8922--8931},
  publisher = {Computer Vision Foundation / {IEEE}},
  year      = {2021},
  url       = {https://openaccess.thecvf.com/content/CVPR2021/html/Sun\_LoFTR\_Detector-Free\_Local\_Feature\_Matching\_With\_Transformers\_CVPR\_2021\_paper.html},
  doi       = {10.1109/CVPR46437.2021.00881},
  bibsource = {dblp computer science bibliography, https://dblp.org}
}

@inproceedings{wang2021ibrnet,
  author    = {Wang, Qianqian and Wang, Zhicheng and Genova, Kyle and Srinivasan, Pratul and Zhou, Howard and Barron, Jonathan T. and Martin-Brualla, Ricardo and Snavely, Noah and Funkhouser, Thomas},
  booktitle = {2021 IEEE/CVF Conference on Computer Vision and Pattern Recognition (CVPR)},
  title     = {{IBRNet}: Learning Multi-View Image-Based Rendering},
  year      = {2021},
  volume    = {},
  number    = {},
  pages     = {4688-4697},
  doi       = {10.1109/CVPR46437.2021.00466}
}

@inproceedings{wewer24latentsplat,
  author    = {Wewer, Christopher
               and Raj, Kevin
               and Ilg, Eddy
               and Schiele, Bernt
               and Lenssen, Jan Eric},
  editor    = {Leonardis, Ale{\v{s}}
               and Ricci, Elisa
               and Roth, Stefan
               and Russakovsky, Olga
               and Sattler, Torsten
               and Varol, G{\"u}l},
  title     = {{latentSplat}: Autoencoding Variational Gaussians for Fast Generalizable {3D} Reconstruction},
  booktitle = {Computer Vision -- ECCV 2024},
  year      = {2025},
  publisher = {Springer Nature Switzerland},
  address   = {Cham},
  pages     = {456--473},
  isbn      = {978-3-031-73021-4}
}

@inproceedings{xu2024depthsplat,
  author    = {Xu, Haofei and Peng, Songyou and Wang, Fangjinhua and Blum, Hermann and Barath, Daniel and Geiger, Andreas and Pollefeys, Marc},
  booktitle = {2025 IEEE/CVF Conference on Computer Vision and Pattern Recognition (CVPR)},
  title     = {{DepthSplat}: Connecting Gaussian Splatting and Depth},
  year      = {2025},
  volume    = {},
  number    = {},
  pages     = {16453-16463},
  doi       = {10.1109/CVPR52734.2025.01534}
}

@inproceedings{yang2023unisim,
  author    = {Yang, Ze and Chen, Yun and Wang, Jingkang and Manivasagam, Sivabalan and Ma, Wei-Chiu and Yang, Anqi Joyce and Urtasun, Raquel},
  booktitle = {2023 IEEE/CVF Conference on Computer Vision and Pattern Recognition (CVPR)},
  title     = {{UniSim}: A Neural Closed-Loop Sensor Simulator},
  year      = {2023},
  volume    = {},
  number    = {},
  pages     = {1389-1399},
  doi       = {10.1109/CVPR52729.2023.00140}
}

@inproceedings{ye2025nopose,
  author    = {Ye, Botao and Liu, Sifei and Xu, Haofei and Li, Xueting and Pollefeys, Marc and Yang, Ming-Hsuan and Peng, Songyou},
  booktitle = {International Conference on Learning Representations},
  editor    = {Y. Yue and A. Garg and N. Peng and F. Sha and R. Yu},
  pages     = {54009--54033},
  title     = {No Pose, No Problem: Surprisingly Simple {3D} Gaussian Splats from Sparse Unposed Images},
  url       = {https://proceedings.iclr.cc/paper_files/paper/2025/file/857b34d81f0a8bfe3e18879dee3b5086-Paper-Conference.pdf},
  volume    = {2025},
  year      = {2025}
}

@inproceedings{yu2021pixelnerf,
  author    = {Yu, Alex and Ye, Vickie and Tancik, Matthew and Kanazawa, Angjoo},
  booktitle = {2021 IEEE/CVF Conference on Computer Vision and Pattern Recognition (CVPR)},
  title     = {{pixelNeRF}: Neural Radiance Fields from One or Few Images},
  year      = {2021},
  volume    = {},
  number    = {},
  pages     = {4576-4585},
  doi       = {10.1109/CVPR46437.2021.00455}
}

@inproceedings{yu2024mip-splatting,
  author    = {Yu, Zehao and Chen, Anpei and Huang, Binbin and Sattler, Torsten and Geiger, Andreas},
  booktitle = {2024 IEEE/CVF Conference on Computer Vision and Pattern Recognition (CVPR)},
  title     = {{Mip-Splatting}: Alias-Free {3D} Gaussian Splatting},
  year      = {2024},
  volume    = {},
  number    = {},
  pages     = {19447-19456},
  doi       = {10.1109/CVPR52733.2024.01839}
}

@inproceedings{zhang2025transplat,
  author    = {Zhang, Chuanrui and Zou, Yingshuang and Li, Zhuoling and Yi, Minmin and Wang, Haoqian},
  title     = {{TranSplat}: generalizable {3D} Gaussian splatting from sparse multi-view images with transformers},
  year      = {2025},
  isbn      = {978-1-57735-897-8},
  publisher = {AAAI Press},
  url       = {https://doi.org/10.1609/aaai.v39i9.33070},
  doi       = {10.1609/aaai.v39i9.33070},
  booktitle = {Proceedings of the Thirty-Ninth AAAI Conference on Artificial Intelligence and Thirty-Seventh Conference on Innovative Applications of Artificial Intelligence and Fifteenth Symposium on Educational Advances in Artificial Intelligence},
  articleno = {1097},
  numpages  = {9},
  series    = {AAAI'25/IAAI'25/EAAI'25}
}

@article{zhou2018stereo,
  author     = {Zhou, Tinghui and Tucker, Richard and Flynn, John and Fyffe, Graham and Snavely, Noah},
  title      = {Stereo magnification: learning view synthesis using multiplane images},
  year       = {2018},
  issue_date = {August 2018},
  publisher  = {Association for Computing Machinery},
  address    = {New York, NY, USA},
  volume     = {37},
  number     = {4},
  issn       = {0730-0301},
  url        = {https://doi.org/10.1145/3197517.3201323},
  doi        = {10.1145/3197517.3201323},
  journal    = {ACM Trans. Graph.},
  month      = jul,
  articleno  = {65},
  numpages   = {12}
}

@inproceedings{sarangi2021deepscaletool,
  author    = {Sarangi, Satyabrata and Baas, Bevan M.},
  title     = {{DeepScaleTool}: A Tool for the Accurate Estimation of Technology Scaling in the Deep-Submicron Era},
  booktitle = {2021 IEEE International Symposium on Circuits and Systems (ISCAS)},
  year      = {2021},
  pages     = {1--5},
  doi       = {10.1109/ISCAS51556.2021.9401196}
}
\balance

\clearpage
\appendix[Artifact Appendix]

\subsection{Abstract}

The SCARF artifact was awarded the Artifacts Available, Artifacts Evaluated
(Functional), and Results Reproduced badges. The v1.0.4 artifact provides the
Python functional and cycle simulators, Feature Similarity Depth Reuse,
Scene-Adaptive Early Sparsification, adapters for TranSplat, MVSplat, and
DepthSplat, Chisel RTL, workflow automation, result validators, and a public
physical-design proxy. The release maps Table~1, Figure~8, Figure~11, and their
supporting records to reproducible commands and machine-readable acceptance
rules. The evaluated scope is stated in Section~G.

\subsection{Artifact check-list (meta-information)}

{\small
\begin{itemize}\itemsep0pt
  \item \textbf{Algorithm:} Feature Similarity Depth Reuse and Scene-Adaptive
  Early Sparsification for semantic depth reuse and pre-materialization
  Gaussian pruning.
  \item \textbf{Program:} Python simulator, Chisel RTL, Docker, and Bash
  automation covering the released accelerator datapath, control, and memory
  system used in evaluation.
  \item \textbf{Compilation:} Python~3.10 and CUDA~12.1. RTL checks use
  JDK~11+, sbt~1.9+, Chisel~6.6, and Verilator~5+.
  \item \textbf{Binary:} No precompiled binary is required. The Docker image
  is built locally from the supplied Dockerfile.
  \item \textbf{Models:} TranSplat, MVSplat, and DepthSplat.
  \item \textbf{Data sets:} Re10K, ACID, and DL3DV for key-result evaluation,
  plus the packaged synthetic fixture for installation validation.
  \item \textbf{Runtime environment:} Ubuntu Linux or the supplied CUDA
  container. Separate classic and DepthSplat profiles isolate their distinct
  PyTorch and CUDA dependencies.
  \item \textbf{Hardware:} An NVIDIA GPU with at least 8~GB VRAM for the quick
  workflow, 12~GB or more for quality and mechanism workflows, and a Jetson
  Orin NX 16~GB in MAXN mode with locked clocks for independent Figure~8
  measurements.
  \item \textbf{Metrics:} PSNR, SSIM, LPIPS, latency, speedup, energy, and
  source-bound RTL cycle counts.
  \item \textbf{Output:} JSON records and provenance manifests that bind the
  source revision, inputs, commands, and output hashes, together with validator
  reports and generated figures.
  \item \textbf{Experiments:} Installation, quality, performance, mechanism
  ablation, and sensitivity workflows.
  \item \textbf{Storage and time:} Approximately 20~GB and 20--60 minutes for
  the first container build and quick setup. Full evaluation requires the
  official data sets and several GPU-hours.
  \item \textbf{License:} Apache-2.0 for the artifact code. Third-party data
  and checkpoints remain under their upstream terms.
  \item \textbf{Archived:}
  \SCARFartifacturl{https://doi.org/10.5281/zenodo.22434931}.
\end{itemize}
}

\subsection{Description}

\subsubsection{How to access}

Download the v1.0.4 source archive from
\SCARFartifacturl{https://doi.org/10.5281/zenodo.22434931}, or clone
\SCARFartifacturl{https://github.com/MAdrid1011/SCARF} with recursive submodules. The
archive includes the quick fixture, Dockerfile, model adapters, RTL,
manifests, frozen mechanism configuration, and reviewer workflows. Official
data sets and model checkpoints are acquired separately by the supplied
scripts under their upstream terms, then recorded in the generated input
manifests. These manifests preserve the data set and checkpoint identities
used during evaluation and bind them to the reviewer workflows described
below.

\subsubsection{Models and data}

The supported model adapters are TranSplat, MVSplat, and DepthSplat. The
checkpoint manifest selects the model-compatible weights for each data set.
The following commands prepare the standard sources:

\begin{small}
\begin{verbatim}
bash data/download_checkpoints.sh \
  --profile all
bash data/download_re10k.sh
\end{verbatim}
\end{small}

The scripts record the selected source and place the resulting files in the
paths used by the adapters. DL3DV preparation and its evaluator terms are
documented in \texttt{data/download\_dl3dv.sh} and
\texttt{artifact/CALIBRATION.md}. Data sets and checkpoints are not
redistributed in the archive.

\subsubsection{Hardware and software}

The quick workflow uses an x86-64 Linux host, Docker with NVIDIA support, and
a CUDA~12.1-compatible GPU. Quality and mechanism workflows run the three
model and data set pairs serially and benefit from at least 12~GB VRAM.
Figure~8 independent measurements use a Jetson Orin NX 16~GB in MAXN mode.
The public RTL path uses JDK~11+, sbt~1.9+, Chisel~6.6, and Verilator~5+ for
elaboration, linting, and simulation.

\subsection{Installation and quick workflow}

The artifact-root commands below build the pinned Functional container and
record its validation output:

\begin{small}
\begin{verbatim}
docker build -t scarf-ae:1.0.4 .
mkdir -p outputs/docker-quick
docker run --rm --gpus all \
  -v "$PWD/outputs/docker-quick:/results" \
  scarf-ae:1.0.4
\end{verbatim}
\end{small}

The container builds the synthetic fixture, obtains the quick checkpoint when
needed, and writes a structured record under \texttt{/results} for provenance
and subsequent validation.

\subsection{Reviewer workflows}

The frozen, calibrated mechanism configuration is released under
\texttt{artifact/calibration/frozen/}. It lets reviewers begin quality
evaluation with model execution without recalibration. The source-bound RTL
timing manifest binds each cycle record to the released RTL and exported model
workload evaluated during AE.

The unified reviewer plan can run all requested targets. Select
\texttt{proxy} when a Jetson Orin NX is unavailable:

\begin{small}
\begin{verbatim}
python scripts/run_reviewer_claims.py \
  --run-config artifact/key_results_run.json \
  --figure8-device proxy \
  --proxy-input \
  /path/to/reviewer-gpu-proxy.json \
  --output-root outputs/ae
\end{verbatim}
\end{small}

\begin{samepage}
For direct workflows, run quality first, then source-bound timing,
mechanisms, and performance:

\begin{small}
\begin{verbatim}
bash scripts/run_ae.sh quality \
  --output-root outputs/ae
bash scripts/run_ae.sh mechanisms \
  --claim-timing-manifest \
  outputs/timing-backend/manifest.json \
  --output-root outputs/ae
bash scripts/run_ae.sh performance \
  --device orin \
  --claim-timing-manifest \
  outputs/timing-backend/manifest.json \
  --output-root outputs/ae
\end{verbatim}
\end{small}
\end{samepage}

The RTL timing and export scripts generate source-bound timing from exported
workloads. The runner binds packed tensors to the released RTL and runs to
completion without an implicit cycle ceiling. The \texttt{--max-cycles}
argument is diagnostic only. The path records positive cycles for ASIC, FSDR,
SAES, combined, and zero-input runs, with provenance bound to released inputs
and RTL.

\subsection{Evaluation and expected results}

Table~1 reports PSNR, SSIM, and LPIPS for every committed model--data set pair,
and Figure~8 reports nine latency pairs and their geometric-mean speedups.
Figure~11 reports no-optimization, FSDR-only, SAES-only, and combined cycle
paths with reuse-rate, \mbox{utilization}, and event counters. For each target,
\texttt{artifact/evaluation\_catalog.json} records the command, required
inputs, evidence class, and acceptance rule enforced by this validator:

\begin{small}
\begin{verbatim}
python scripts/validate_ae.py \
  --input outputs/ae --require-key-results
\end{verbatim}
\end{small}

Quality tolerances are 0.15~dB PSNR, 0.005 SSIM, and 0.005 LPIPS. Figure~8
and Figure~11 geometric means allow 5\% relative variation. Mechanism rates
and utilization allow two percentage points, while event counts allow 5\%
relative variation against the frozen reference records.

\subsection{Hardware proxy and scope}

Reviewers without Orin hardware can use the portability proxy after capturing
same-workload CUDA timings. The proxy reports labeled latency, FPS, and
speedup estimates without independent Orin power or temperature evidence. The
released public hardware path routes the RTL with iFlow and
ASAP7, then reports the technology-normalized estimate defined by the paper,
with raw and scaled fields kept separate. Commercial TSMC 28~nm and LPDDR
collateral are outside the archive. The Results Reproduced badge applies to
the evidence evaluated during AE within these stated scope boundaries, not to
measurements that depend on proprietary technology files outside the archive.

\end{document}